\documentclass[epj]{svjour}
\usepackage{graphicx}
\usepackage{dcolumn}
\usepackage{bm}
\usepackage{booktabs}
\usepackage[style=numeric,sorting=none]{biblatex} 
\usepackage[caption=false]{subfig}
\def \bstar {$\beta^*$ }

\newcommand{\bs}{$\beta^*$\xspace}

\newcommand{\sig}{$\sigma$\xspace}
\newcommand{\xco}{x_\mathrm{co}}
\newcommand{\dal}{\Delta_\mathrm{ba}}
\newcommand{\dba}{\Delta_\mathrm{ba}}
\newcommand{\dcb}{\Delta_\mathrm{cb}}
\newcommand{\deltap}{\delta_p}
\newcommand{\bbeat}{k_\beta}

\newcommand{\en}{$\epsilon_\mathrm{n}$\xspace}
\newcommand{\madx}{MAD-X\xspace}
\usepackage{xspace}
\usepackage{float}

\usepackage{url}
\usepackage{booktabs}
\usepackage{multirow}
\usepackage{siunitx}
\usepackage{lastpage}
\begin{document}

\title{Beam-based aperture measurements with movable collimator jaws as performance booster of the CERN Large Hadron Collider}

  

\author{N.~Fuster-Mart\'inez\inst{1,2}, R.~W.~A\ss mann\inst{3}, R.~Bruce\inst{2}, M.~Giovannozzi\inst{2}, P.~Hermes\inst{2}, A.~Mereghetti\inst{2}, D.~Mirarchi\inst{2}, S.~Redaelli\inst{2} \and J.~Wenninger\inst{2}
}

%
%
\institute{Instituto de F\'isica Corpuscular (CSIC-UV), Valencia, Spain \and CERN, Geneva, Switzerland \and DESY, Hamburg, Germany}

\date{Received: date / Revised version: date}
%
\abstract{
The beam aperture of a particle accelerator defines the clearance available for the circulating beams and is a parameter of paramount importance for the accelerator performance. At the CERN Large Hadron Collider (LHC), the knowledge and control of the available aperture is crucial because the nominal proton beams carry an energy of 362 MJ stored in a superconducting environment. Even a tiny fraction of beam losses could quench the superconducting magnets or cause severe material damage. Furthermore, in a circular collider, the performance in terms of peak luminosity depends to a large extent on the aperture of the inner triplet quadrupoles, which are used to focus the beams at the interaction points. In the LHC, this aperture represents the smallest aperture at top-energy with squeezed beams and determines the maximum potential reach of the peak luminosity. Beam-based aperture measurements in these conditions are difficult and challenging. In this paper, we present different methods that have been developed over the years for precise beam-based aperture measurements in the LHC, highlighting applications and results that contributed to boost the operational LHC performance in Run~1 (2010-2013) and Run~2 (2015-2018).
}
\maketitle
\section{\label{sec:level1}Introduction}

In most storage rings, a detailed knowledge of the geometric aperture is required in order to guarantee a sufficient beam clearance and a safe operation, to minimise the downtime and optimise the beam losses around the ring. In the CERN Large Hadron Collider (LHC)~\cite{lhcdesignV1}, where proton and heavy-ion beams are brought in collisions for high-energy physics experiments, the control of the available aperture becomes critical due to the nominal stored proton-beam energy of 362 MJ in a superconducting (SC) accelerator environment. Even small losses of the order of tens of $\rm{mJ/cm^3}$ can cause the SC magnets to quench, i.e. changing their state from SC to normal conducting (NC), or in the extreme case material damage that can be extremely costly. 

In this article, we use the term aperture to define the \textit{normalised} aperture at a generic location $s$ along the LHC circumference, $A_{x,y}(s)$ and it is defined as the smallest transverse distance in the transverse $x-y$-plane between the beam centre and the mechanical aperture, $r_{x,y}(s)$, expressed in units of the local transverse beam size, $\sigma_{x,y}$, as
\begin{equation}
    A_{x,y}(s)=r_{x,y}(s)/\sigma_{x,y}(s)\, ,
\end{equation}
where the beam size is modulated with the $\beta$-function as
\begin{equation}
    \sigma_{x,y}(s)=\sqrt{\frac{\beta_{x,y} \epsilon_{N}^\mathrm{design}}{\gamma}} \, .
\end{equation}
Note that here we include the design normalised geometrical emittance, $\epsilon_{N}^\mathrm{design}$ and $\gamma$ is the relativistic factor. This aperture is reduced by a number of imperfections and tolerances, for example, misalignment of magnets and vacuum pipes, orbit errors within machine elements and off-energy offsets, to name a few. This normalised aperture is directly connected to the risk of local beam losses that could potentially limit the LHC performance. 

\begin{figure}[!htbp]
    \centering
    \includegraphics[width=10cm]{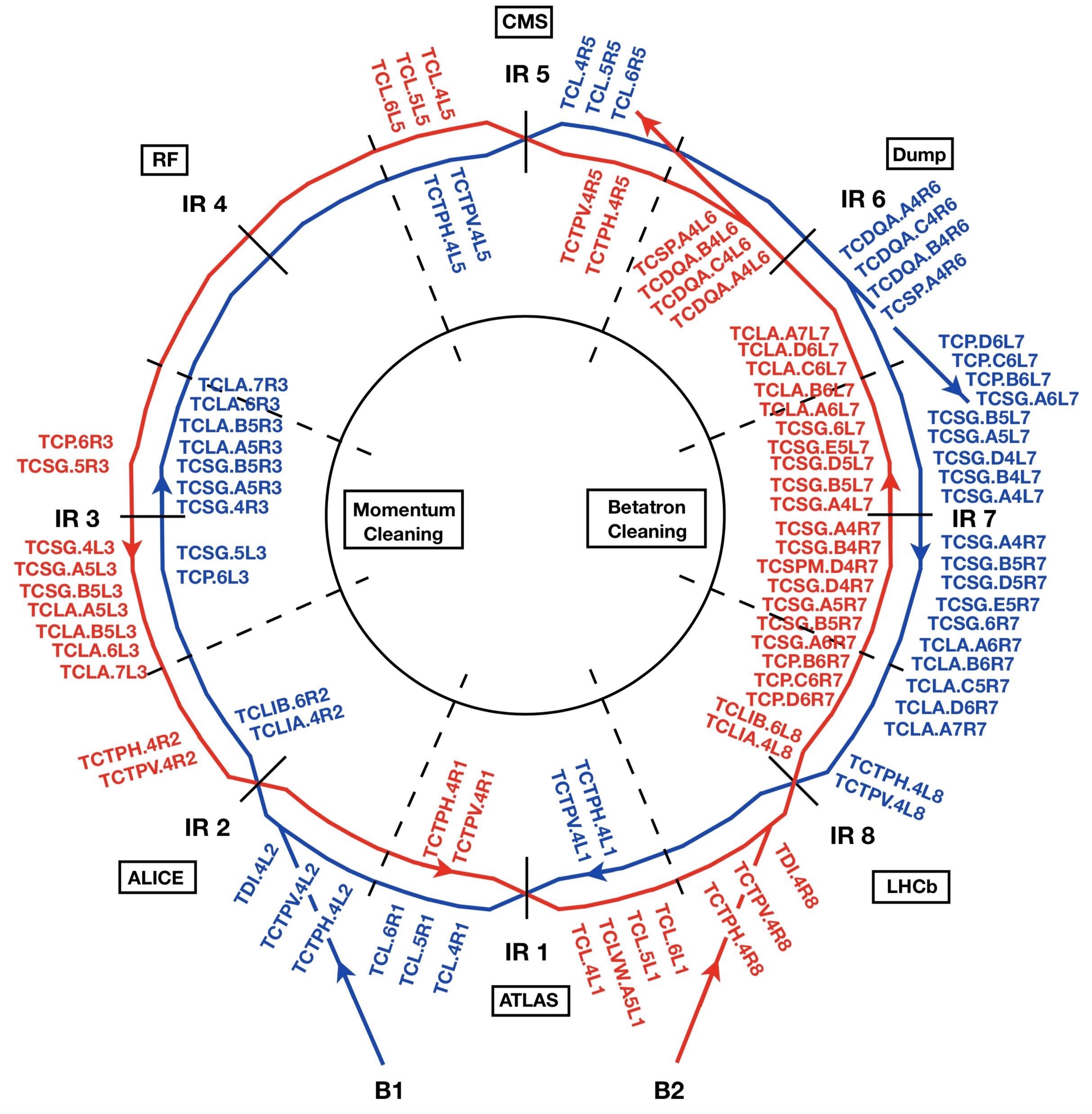}
    \caption{2018 LHC collimation system layout~\cite{Gabi:Crosstalk} in blue and red for B1 and B2, respectively. The names and locations of the different IRs and particle physics experiments are also depicted.}
    \label{LHClayout} 
\end{figure}

A detailed knowledge of the smallest aperture in the ring is very important for maximising the luminosity by decreasing \bs, the $\beta$-function at the interaction points (IPs), to smaller values.  At top-energy, as \bs is decreased, the aperture of the inner triplet quadrupoles used for the final focus of the beams at the IPs decreases, as the local beam size increases, and such a decrease is only allowed down to the level that can be protected by the collimation system. It is typically the smallest aperture in the machine, known as the \textit{global} aperture bottleneck, that determines the minimum \bs reach, which is proportional to the peak luminosity. The LHC has eight Interaction Regions (IRs) of which four are dedicated for the particle physics experiments ATLAS (IR1), ALICE (IR2), CMS (IR5) and LHCb (IR8). ATLAS and CMS are the general-purpose, high-luminosity experiments, whereas ALICE and LHCb are specialised experiments, operating at lower luminosity levels. In these IRs, the two counter-rotating LHC beams named as Beam~1 (B1, the clockwise beam) and Beam~2 (B2, the counter-clockwise beam) collide. Fig.~\ref{LHClayout} shows the layout of the LHC collimation system along with the trajectory of the beams; B1 in blue and B2 in red.

\begin{figure}[!htbp]
    \centering
    \includegraphics*[width=8cm]{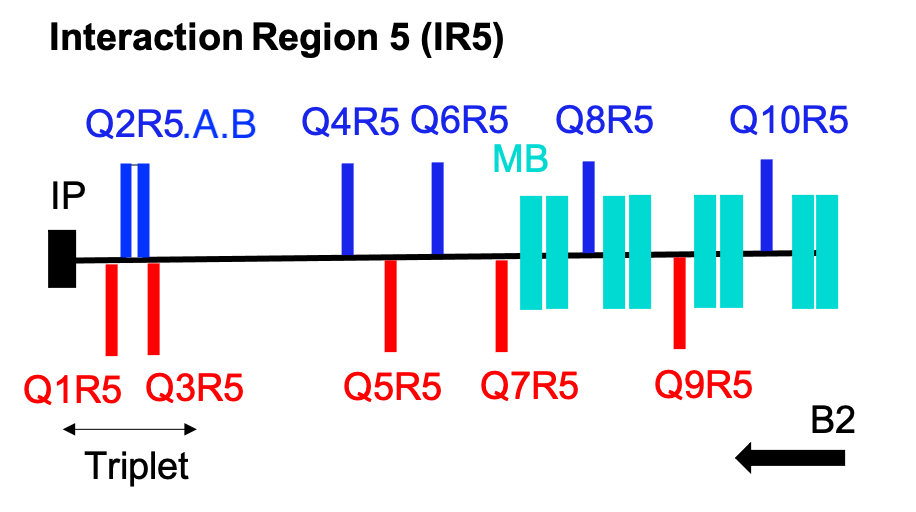}
    \caption{Schematic layout of an IP with main focusing (blue) and defocusing (red) quadrupoles and the dipoles (light blue) from the IP to Q10 illustrating the naming convention used in this paper. Here illustrated for B2 to the right of IR5.}
    \label{IPscheme} 
\end{figure}

In this paper, in order to refer to a certain quadrupole in relation to an IR, the name is composed of the number identifying the position with respect to the IP, the side with respect to the IP, left (L) or right (R), and a number defining the corresponding IR. Fig.~\ref{IPscheme} illustrates this convention for the case of the right side of IR5. In this figure the layout until the dispersion suppressor is depicted in which the main quadrupoles (Q) and dipoles (MB) are indicated. The final focus triplet magnets correspond to Q1R5, Q2R5 and Q3R5.

The settings of the LHC multi-stage collimation system~\cite{assmann05chamonix,guillaume-thesis,assmann06,chiara-thesis,bruce14_PRSTAB_sixtr,valentino17_PRSTAB}, designed to protect the LHC from normal and abnormal losses, define the smallest normalised aperture that can be protected. The available aperture together with the collimator settings impose limitations on the achievable \bs and thus on the luminosity reach~\cite{bruce15_PRSTAB_betaStar,bruce17_NIM_beta40cm}. This implies that detailed aperture knowledge and efficient beam-based techniques to probe the available aperture in the machine are key elements to boost the collider's performance, which is indeed the case for the LHC. In fact, together with the optimisation of the settings of the collimation system and optics~\cite{bruce17_NIM_beta40cm}, the detailed knowledge of the LHC aperture has allowed the \bs at the high-luminosity experiments to be decreased throughout Run 1 and Run 2 as illustrated in Table
~\ref{beta*}. In 2018, the LHC operated with \bs = 25~cm in IR1/5 (as can be seen in Table~\ref{beta*}), representing a huge gain compared to the nominal LHC operational scenario with \bs =  55~cm, and even more with respect to the initial LHC configuration from 2010 of \bs = 3.5~m. In Run~2, decreasing \bs has proven to be a very efficient way of increasing the peak performance even in presence of limitations on the beam intensity.

\begin{table}[!htbp]
  \begin{center}
 
    \caption{Run 1 and Run 2 minimum \bs values reached in each LHC particle physics experiment IRs and beam energy per year.}
    \begin{tabular}{lccccccccc}
      \hline
      \multirow{2}{*}{}&   \multicolumn{7}{c}{\textbf{Year}}\\
      &2010&2011&2012&2015&2016&2017&2018\\\hline\hline
     $\beta^{*}$ IP1/5 [cm]&  350     & 150-100 &60&80&40&30&25\\ 
     $\beta^{*}$ IP2 [cm]&    3.5    & 10 & 3 &10&10&10&10\\ 
     $\beta^{*}$ IP8 [cm]&    3.5    & 3 &3 &3&3&3&3\\
     E [TeV]&3.5&4&4&6.5&6.5&6.5&6.5\\\hline
    
    \end{tabular}
    \label{beta*}
  \end{center}
\end{table}

\begin{figure}[!htbp]
    \centering
    \includegraphics[width=14cm]{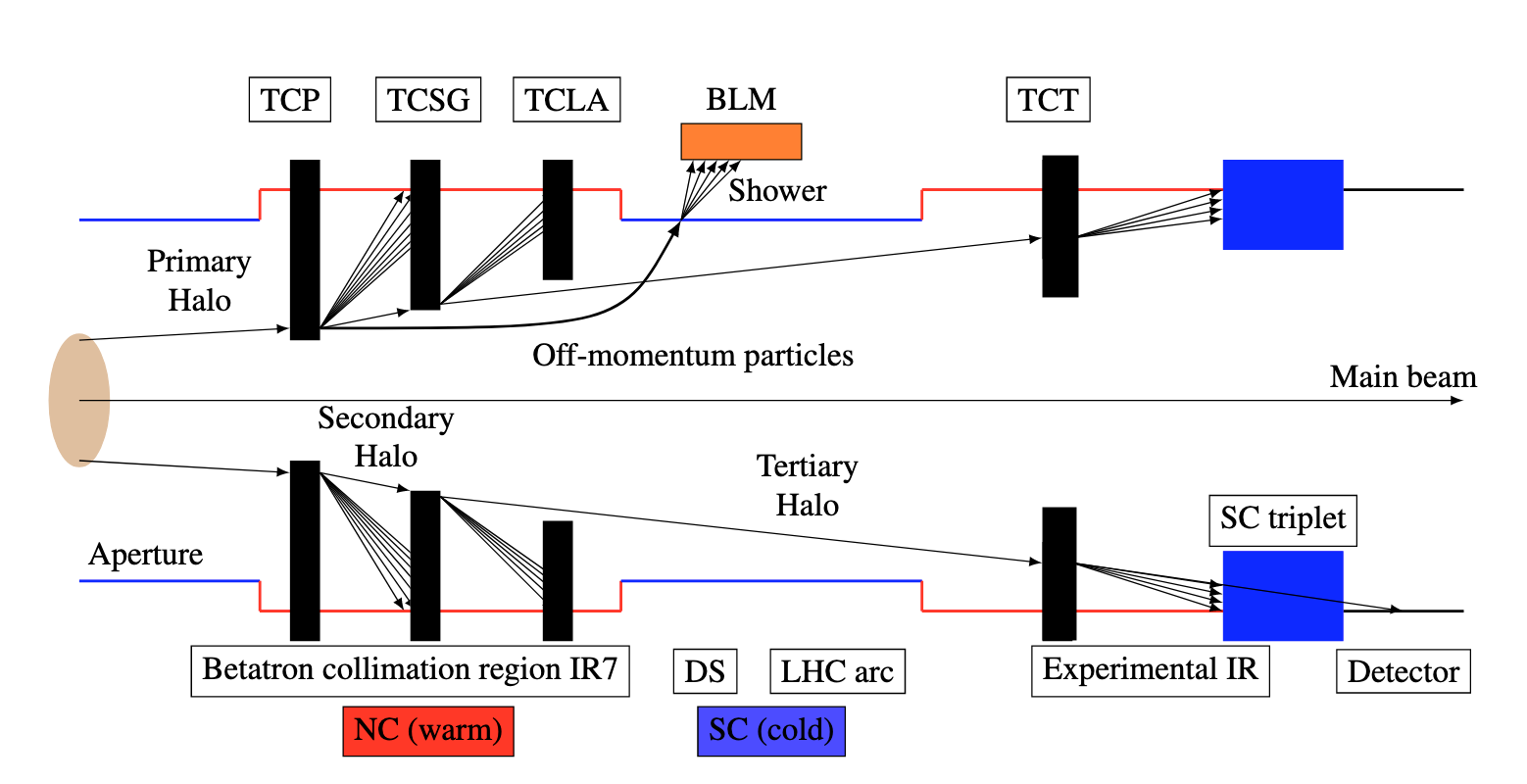}
    \caption{Sketch of the LHC multi-stage collimation system in IR7 and on the experimental IRs, where the different collimator families are indicated. NC and SC apertures are depicted in red and blue, respectively~\cite{Redaelli:2646800}.}
    \label{Coll_scheme} 
\end{figure}

The LHC multi-stage collimation system is organised in a well-defined transverse hierarchy, as illustrated in Fig~\ref{Coll_scheme}, with different collimator families where each individual device consists of two movable jaws with the beam passing through the centre. Primary collimators (TCPs) made of carbon-fiber-composite (CFC) are the closest collimators to the beams. After the TCPs, secondary collimators (TCSGs) made of CFC and active absorbers (TCLAs) made of Inermet-180 (heavy tungsten alloy) are placed to absorb particles out-scattered by the TCPs and TCSGs (secondary and tertiary beam halo, respectively) as illustrated in Fig.~\ref{Coll_scheme}. 

These collimators are located in two dedicated IRs in the LHC: IR7 for betatron cleaning and IR3 for off-momentum cleaning (see Fig.~\ref{LHClayout}). The tertiary collimators (TCTs) made of Inermet-180 are installed upstream of the particle physics experiments in IR1, IR2, IR5 and IR8 (see Fig.~\ref{LHClayout}). These collimators are there to absorb the tertiary betatron beam halo and provide passive protection of the aperture bottlenecks in the triplet quadrupoles of the final focusing system, as well as to keep a good control of the experimental backgrounds~\cite{bruce13_NIM_backgrounds,bruce19_PRAB_beam-halo_backgrounds_ATLAS}. Downstream of the experiments, other collimators are installed to absorb the debris from the particle physics collisions (TCLs). Finally, two more CFC collimators per beam (TCSP and TCDQ) are installed in the dump region (IR6) for beam-dump protection. These collimators must ensure the protection of the machine in case of beam dump failures~\cite{assmann02dump,schmidt06,bruce15_PRSTAB_betaStar,bruce17_NIM_beta40cm}. All collimator jaws are 1~m long, except for TCPs and TCDQ that are 60~cm and 9~m long, respectively. The detailed 2018 LHC multi-stage collimation system layout is depicted in Fig.~\ref{LHClayout}.

This paper is organised as follows. In Section~\ref{sec:level2}, the methods used to estimate the beam aperture are described in detail. In Section~\ref{sec:level3}, all beam-based aperture measurement techniques developed for the LHC during Run~1 and Run~2 are presented. In Section~\ref{sec:level4}, a summary of the global aperture measurements performed in the LHC during Run~1 and Run~2 at injection energy is presented, as well as a benchmark comparison study between the methods. Additionally, local and global aperture measurements carried out at top-energy for \bs reach studies in the high-luminosity experiments, ATLAS and CMS, performed during Run~1 and Run~2 are presented and compared with expectations.

\section{\label{sec:level2}Aperture model}

In circular particle accelerators, the concept of machine aperture goes beyond the simple definition of the cross-section of the mechanical design of the beam pipes. It also depends on beam properties,  errors (on beam or mechanical properties), and on the machine configuration (optics choice, orbit settings, energy value, etc.). For evaluating the aperture accurately, the following aspects must be taken into account: the transverse mechanical dimensions of the machine components; the alignment errors of the machine elements; the beam orbit; the beam optics, which influence the beam size; the magnetic-field errors, which affect the beam orbit and optics; and the collimator settings that determine the size and population of beam halo. All these quantities define an aperture model that is refined by measurements in order to calculate the global aperture bottleneck of the machine.

Two different models to estimate the apertures have been used in the LHC. The first one, called the \textit{$n_1$} model~\cite{note66,note111,lhcdesignV1}, was used from the design phase of the LHC. It calculates the most limiting aperture using the realistic transverse geometry and taking into account different possible errors that are added linearly, i.e. in a conservative way. The model was updated at the end of Run~1 by adding realistic tolerances matching the measured apertures. A second model was developed thanks to the improved accuracy of the beam-based aperture measurement techniques, based on the scaling of the aperture measured in previous years and taking into account the optics and orbit parameters at the location of the bottleneck. These two models are described in detail in this section.

\subsection{The 2D LHC aperture model}
The $n_1$ aperture model, implemented in \madx and described in detail in Refs.~\cite{note66,note111,lhcdesignV1}, calculates the most limiting aperture in the transverse 2D plane, through a numerical scan over different azimuthal angles considering a real aperture geometry. Before performing the scan, the initial displacement of the beam centre is calculated as the vector sum of the nominal closed orbit, the closed orbit deviation $\xco$, the mechanical alignment tolerance $\dal$, the beam-screen alignment $\dba$, the cold-bore alignment $\dcb$, as well as an off-momentum component $\tilde{D}\deltap$. It should be noted that the scan is performed over the azimuthal angle originating at $\xco$, so as to catch possible restrictions in the skew plane, and that the dispersion contribution $\tilde{D}$, multiplying the momentum offset $\deltap$ at a given location $s$ in the ring, is given in each plane by 
\begin{equation}
\tilde{D}(s) = k_\beta \left(|D(s)| + D_\mathrm{arc}\, f_\mathrm{arc}\, \sqrt{\frac{\beta(s)}{\beta_\mathrm{arc}}}\right),  
\end{equation}
where $k_\beta$ is the change of the beam size and dispersion due to $\beta$-beating, $D_\mathrm{arc}$ is the peak dispersion in the arc, $f_\mathrm{arc}$ is the fractional parasitic dispersion, and $\beta_\mathrm{arc}$ is the optical $\beta$-function in a focusing arc quadrupole. The linear sum of these errors is motivated by statistical considerations, i.e. given the large number of SC magnets (e.g. 1232 arc dipoles), as it is not unlikely that somewhere all errors add coherently. After this calculation, the scan is started and the beam extent is scaled until it reaches the physical aperture at some of the azimuthal angles. The resulting beam extent is then normalised by the local beam size. At the LHC, the design value $\epsilon_\mathrm{n}=3.5~\mu$m is used to calculate the beam size at a given location, in order to have comparable results, even if the real value of $\epsilon_\mathrm{n}$ can vary between fills. This normalisation is motivated, as we will see later, by the fact that in the measurements the aperture is compared to the opening of the jaws of a reference collimator, computed using the design $\epsilon_\mathrm{n}$. In addition, the resulting beam size value is multiplied by $\bbeat$ to account for the change of the beam size due to the $\beta$-beating given as an input parameter for the calculations.

During the LHC Run~1, several potential errors affecting the $n_1$ model were found to be less restrictive than in the worst-case scenarios assumed during the design phase of the LHC. This was the case for the orbit control, optics correction, and the alignment of the machine elements, to name a few. Because of that, the tolerances for the $n_1$ calculations had to be adjusted. Table~\ref{n1_param} summarises the MAD-X aperture module input parameters used to calculate the LHC aperture model at top-energy with squeezed beams, in the design phase and later based on the measurements in operational conditions in 2012 for the proton run at 4 TeV~\cite{bruce14_n1_ap_meas}. 

\begin{table}[!htbp]
\caption{Input parameters used to calculate the LHC ring aperture bottleneck in units of \sig using the $n_1$ model. Two sets of input are presented: the design-phase LHC parameters, and those measured~\cite{ipac11_assmann_aperture, ipac12_redaelli_aperture} in operational conditions in 2012.
\label{n1_param}}
\centering
\begin{tabular}{lcc}\hline
\textbf{Parameter} &\textbf{LHC} &\textbf{LHC}\\
&\textbf{design}&\textbf{measured}\\\hline\hline
Normalised emittance [$\mu$m] &   3.75&3.5  \\
Radial closed orbit excursion, $x_{co}$ [mm] &  3  &0.5  \\
Momentum offset, $\delta_p$ &8.6$\times$10$^{-4}$&2$\times$10$^{-4}$ \\
$\beta$-beating fractional $\sigma$ change, $\kappa_{\beta}$   & 1.1&1.025  \\
Relative parasitic dispersion, $f_\mathrm{arc}$  & 0.27&0.1  \\\hline
\end{tabular}
\end{table}

\begin{figure}[!htbp]
   \centering
   \includegraphics[width=8cm]{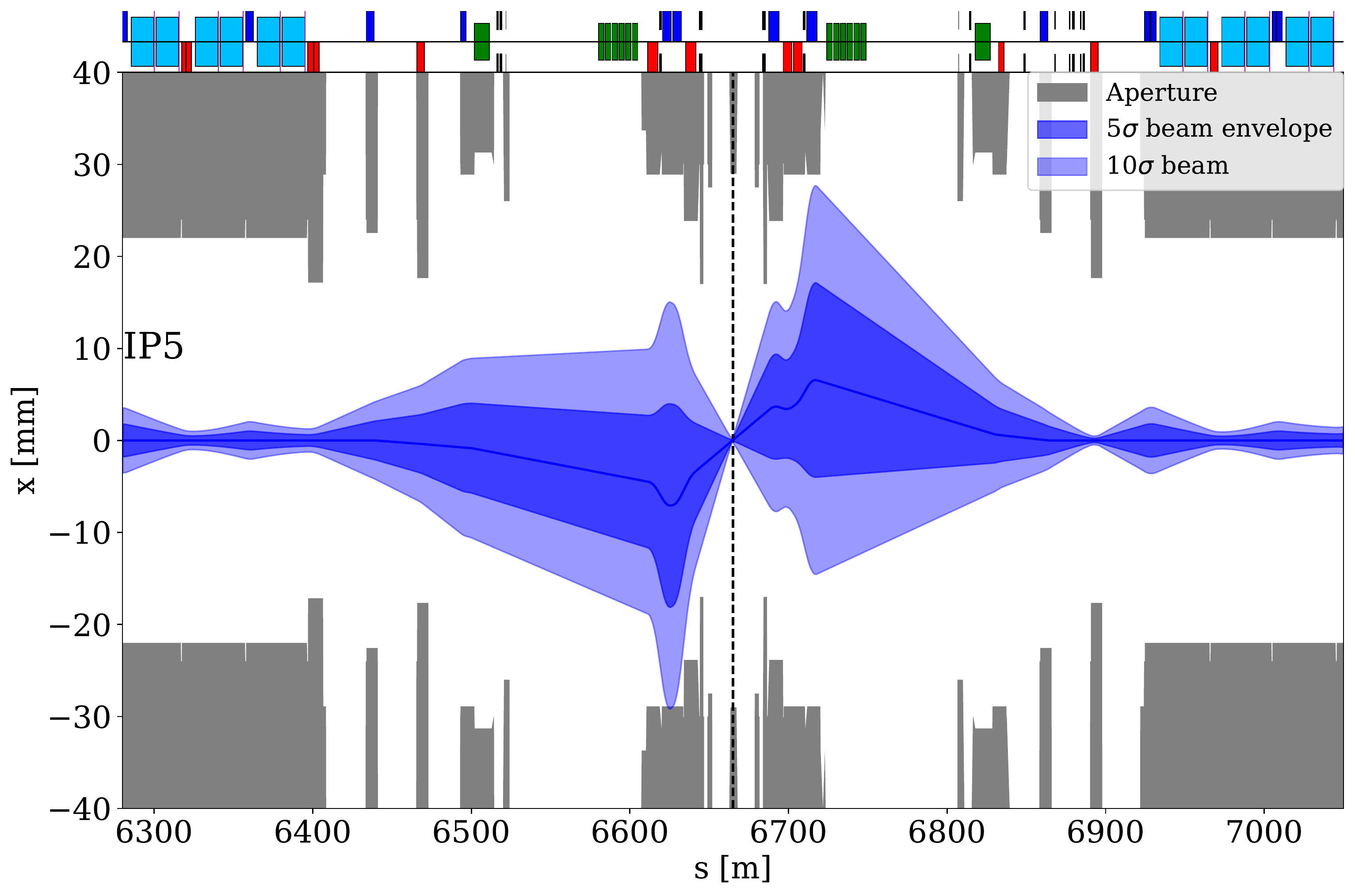}
    \includegraphics[width=8cm]{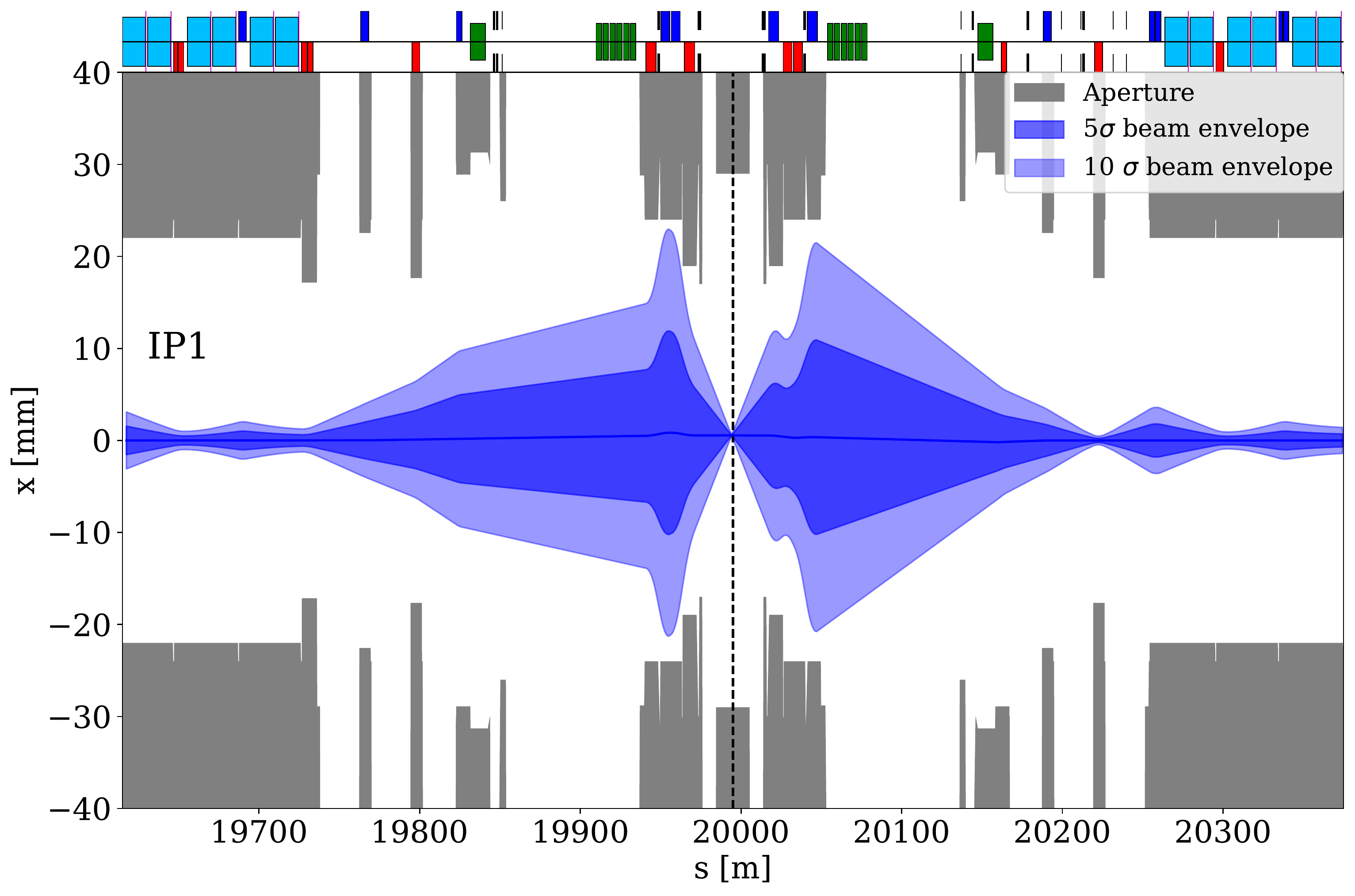}
         \includegraphics[width=8cm]{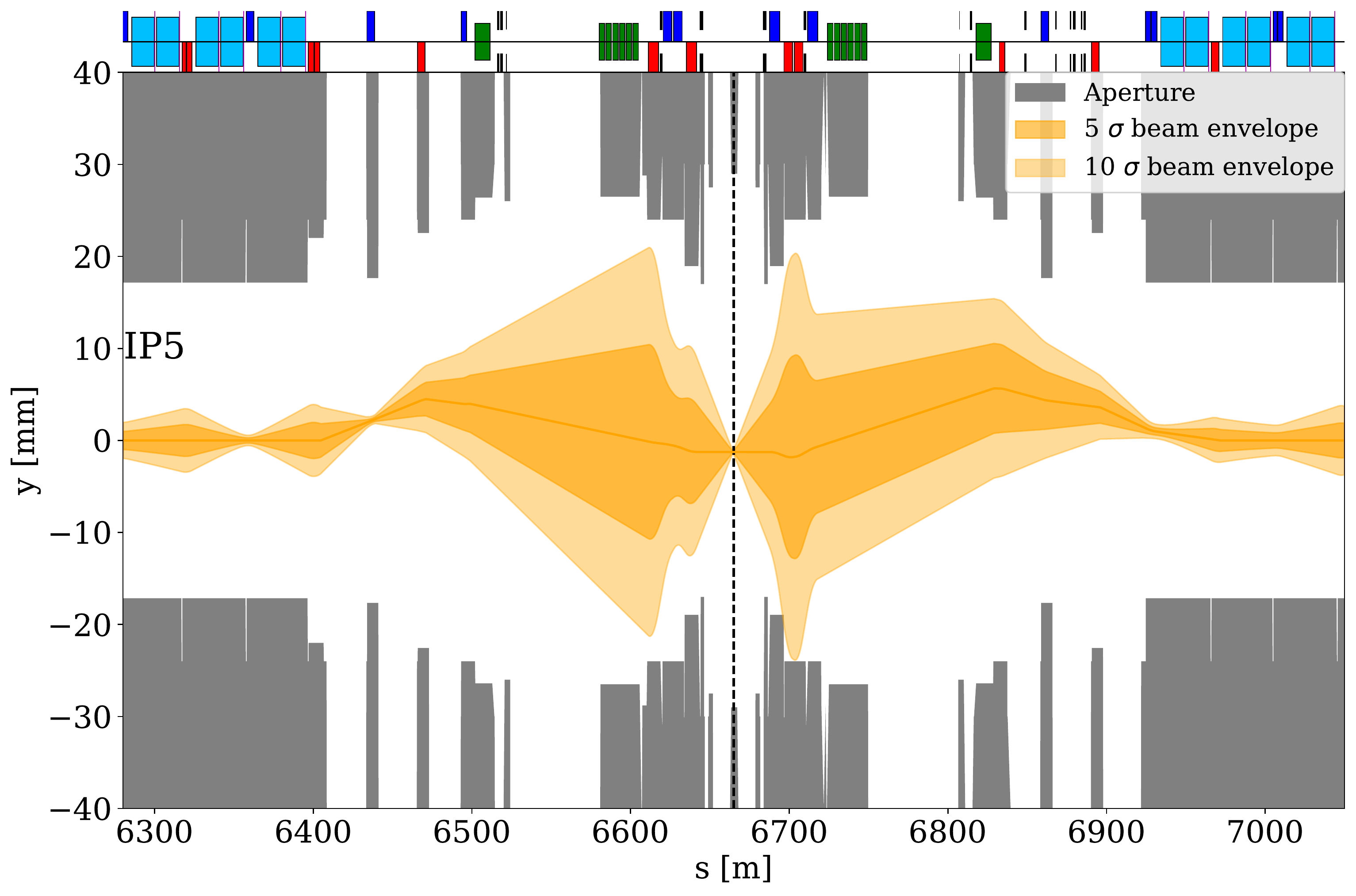}
    \includegraphics[width=8cm]{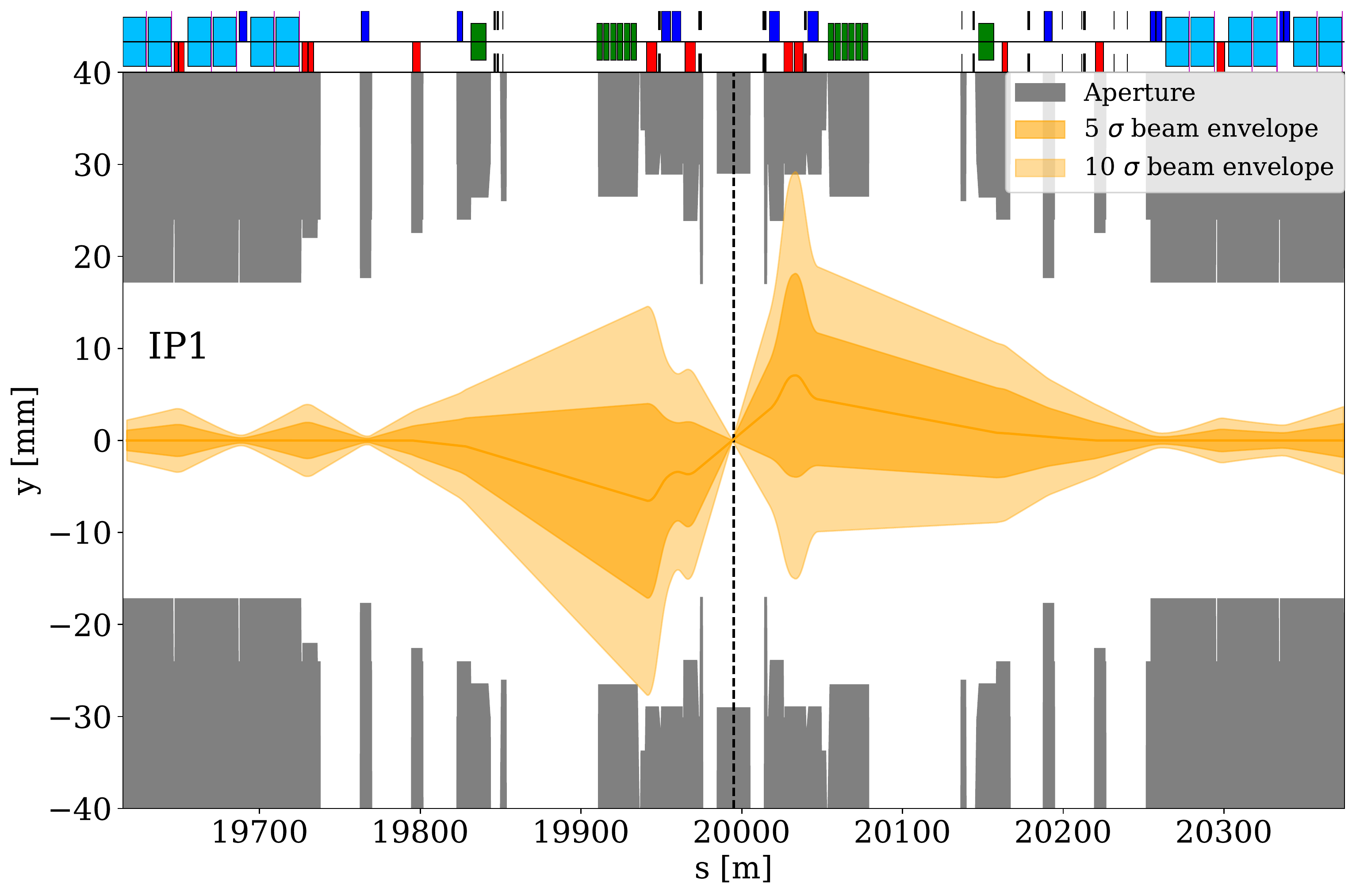}
      \includegraphics[width=8cm]{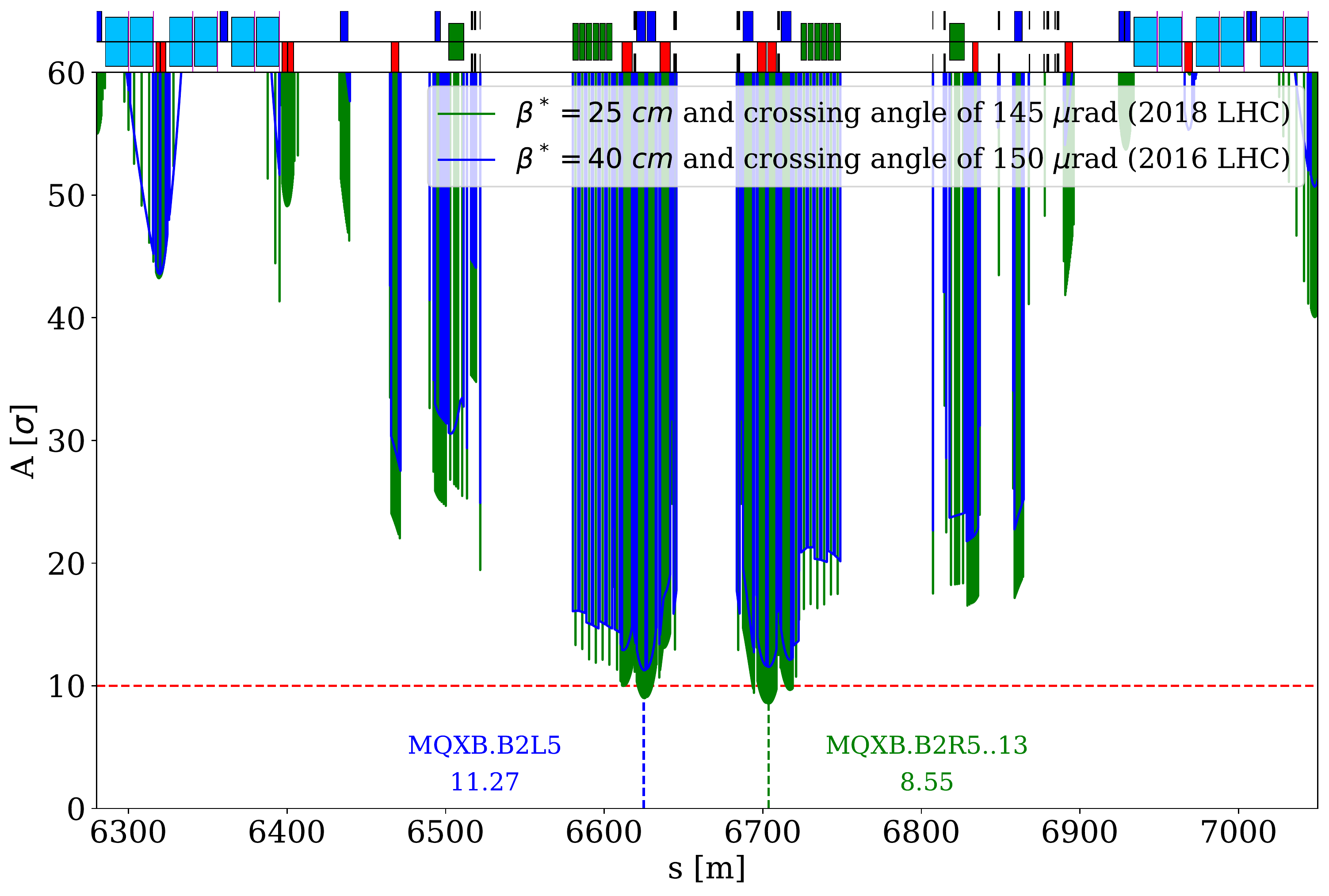}
            \includegraphics[width=8cm]{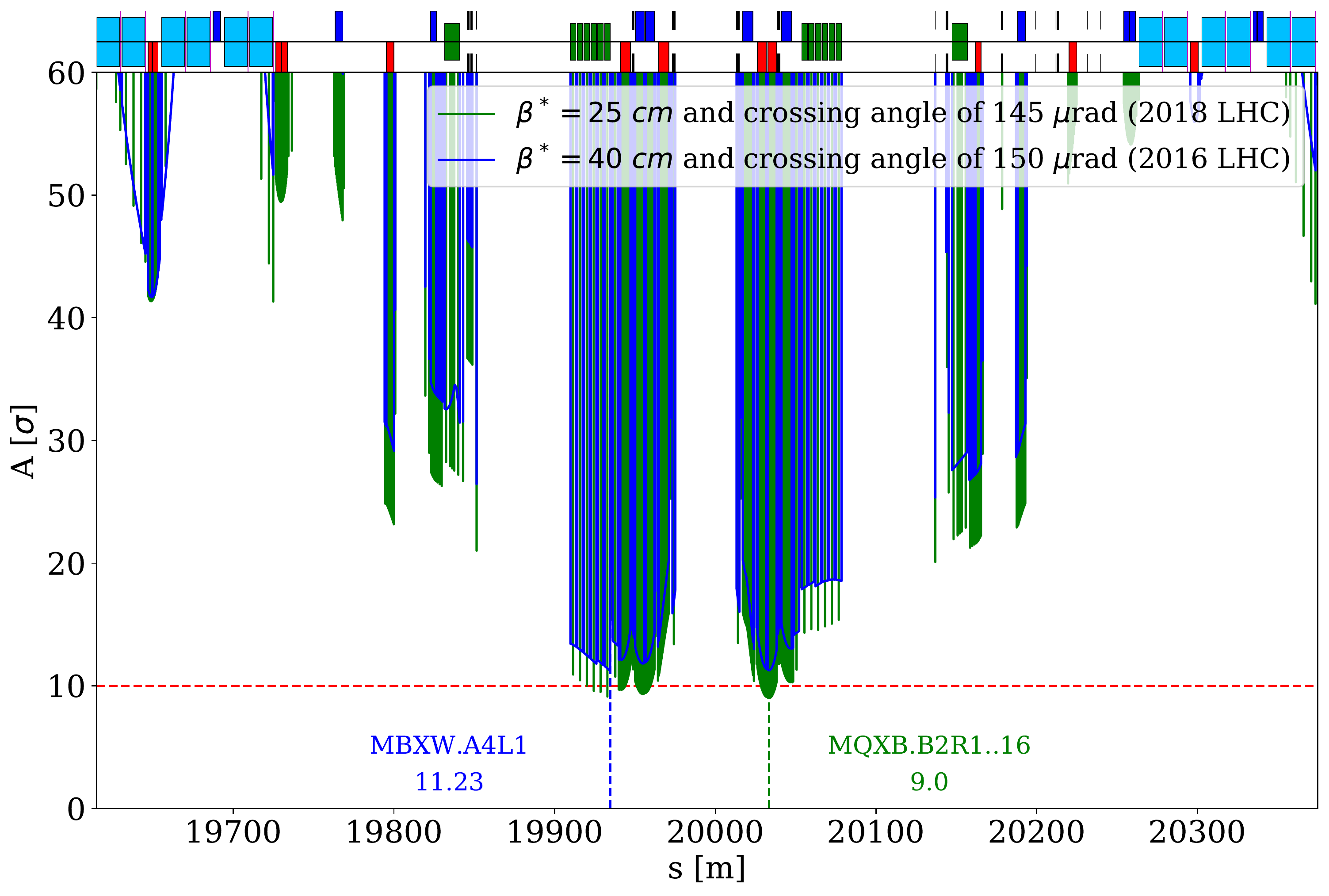}
   \caption{Horizontal (top) and vertical (middle) beam envelopes of 5 and 10~\sig for the 2018 optics with a \bs of 25~cm and a half crossing angle of 145~$\mu$rad, centred around the closed orbit, in IR5 (left) and IR1 (right). The geometrical aperture model is also plotted, and the machine layout is depicted on the top of the plots with dipoles in light blue, focusing quadrupoles in blue, defocusing quadrupoles in red, sextupoles in green and collimators in black. The bottom plots show an example of the aperture calculated in units of \sig with the $n_1$ model for two different LHC optics configurations in IR5 (left) and IR1 (right) using the measured parameters from Table~ \ref{n1_param}. The aperture results are shown for the 2018 optics in green and for the 2016 optics in blue. The bottleneck location and corresponding aperture are also indicated as well as a red dashed line at 10 $\sigma$ for reference.}
   \label{n1_example}
\end{figure}

At top-energy and with squeezed beams the bottleneck is expected, and was always measured in Run~1 and Run~2, to be in the inner triplet quadrupoles in the high-luminosity experiments (IR1 and IR5), where the smallest \bs is deployed for proton runs. In order to reduce the beam-beam interactions and maximise the luminosity and beam stability, the LHC beams collide with an optimised crossing angle. The plane in which this angle is introduced is called the crossing plane. The plane perpendicular to the crossing plane is called the separation plane, as in this plane the beams have an offset until the separation bump is collapsed to bring the beams into collision. For large crossing angles, the bottleneck was always found to be in the crossing plane; however, for reduced crossing angle values, the limiting aperture in both the crossing and the separation planes are similar, as is illustrated in Fig.~\ref{n1_example} (top and middle plots) for the 2018 collision optics with \bs=25~cm and a half crossing angle of 145~$\mu$rad. The top and middle plots in Fig.~\ref{n1_example} show the horizontal and vertical closed orbit in IR5 (left) and IR1 (right) with a beam envelope of 5 and 10 \sig. The mechanical aperture model and the machine layout are also depicted. The crossing angle is visible on the horizontal closed orbit in IP1 and on the vertical closed orbit in IP5. The bottom plots of Fig.~\ref{n1_example} show the resulting $n_1$ aperture model in units of \sig for two different LHC optical configurations, 2018 (in green) and for the 2016 (in blue), in IR5 (left) and IR1 (right). The measured parameters from Table~ \ref{n1_param} have been used in this calculation. The bottleneck location and corresponding aperture in units of \sig are also indicated for each optics configuration. A smaller aperture by 2~\sig can be observed for the \bs=25~cm with respect to a \bs=40~cm. In Section~\ref{sec:level4} these calculations are compared with the measured values.
\subsection{Aperture-scaling model}
The aperture-scaling model was developed in order to make fast predictions of the local aperture of the triplets in the experimental IRs at different values of \bs and crossing angle, starting from a known aperture obtained through beam-based measurements using any of the techniques described later. The immediate application was to evaluate the \bs-reach due to the reduced aperture in the triplets~\cite{bruce15_PRSTAB_betaStar}.   

In general, it is not possible to calculate the aperture for a given machine configuration using data acquired using a different one (e.g. different optics, orbit etc.), since the location of the limiting aperture might move in both the longitudinal and transverse planes. However, whenever the location is expected to stay unchanged, we can use simple scaling laws to estimate the aperture for a different configuration: this turned to be a good approximation in the case of the triplets~\cite{bruce10_evian}. 

The aperture-scaling method consists of using the measured aperture $n_\mathrm{m}$ in units of local beam \sig for a given optics configuration with subscript $\mathrm{m}$, as a starting point. For another configuration (denoted by subscript $\mathrm{sc}$), the aperture $n_\mathrm{sc}$ can be computed by accounting for the change in the $\beta$-function and the orbit at the location of the bottleneck. Since the physical aperture is identical in the two configurations, it must hold that 
\begin{equation}
\label{eq:nm}
    n_\mathrm{m} \sigma_{u\mathrm{m}} + |u_\mathrm{m}| = n_\mathrm{sc} \sigma_{u\mathrm{sc}} + |u_\mathrm{sc}| \, ,
\end{equation}
where $u$ is the transverse coordinate of the orbit in the limiting plane (we assume here that the bottleneck lies purely in the horizontal or vertical plane and we use the absolute value of $u$ in order to account for cases where the orbit is negative), and $\sigma_u\approx\sqrt{\beta_u \epsilon_\mathrm{n}/\gamma}$ is the local beam size in the $u$-plane at a small-dispersion location at high energy, where $\beta_u$ is the local $\beta$-function, $\epsilon_\mathrm{n}$ the normalised transverse emittance, and $\gamma$ the relativistic factor.  

The ideal closed orbit $u$ and $\beta$-function can be extracted from optics calculations in both configurations. To account for the fact that the actual machine tolerances, we also include an additional orbit shift $\delta u$ between the two studied configurations, as well as $\beta$-beating factors $\lambda$. Solving Eq.~(\ref{eq:nm}) for $n_\mathrm{sc}$, and including the tolerances $\delta u$ and $\lambda$, we obtain
\begin{equation}
    n_\mathrm{sc} = \frac{|u_\mathrm{m}|-|u_\mathrm{sc}|-\delta u}{\sqrt{\beta_{u\mathrm{sc}} \lambda_\mathrm{sc} \epsilon_\mathrm{n}/\gamma_\mathrm{sc}}} + n_\mathrm{m} \sqrt{ \frac{\lambda_\mathrm{m} \beta_{u\mathrm{m}}\gamma_\mathrm{sc}}{\lambda_\mathrm{sc} \beta_{u\mathrm{sc}}\gamma_\mathrm{m}}} \, .
\end{equation}

Experience has shown that the orbit tolerance $\delta u$ can usually be omitted and that a ratio  $\lambda_\mathrm{m}/\lambda_\mathrm{sc}$ of a few percent is reasonable, given the obtained level of optics correction in the LHC~\cite{tomas09prstab}.

The method has been shown to provide reliable predictions of the local aperture bottleneck as long as the aperture limit lies clearly in one of the two transverse planes, effectively making the aperture calculation 1D, and that there are no major optics or orbit differences that could shift the location of the aperture bottleneck. It has been shown to be both in excellent agreement with the $n_1$ model if realistic tolerances are considered~\cite{bruce14_evian} and with beam-based aperture measurements~\cite{bruce17_NIM_beta40cm}. This method has the advantage of being very fast and removes some uncertainties in the error tolerances, since it starts from the outcome of beam-based measurement with a precision higher than those uncertainties, as will be illustrated in Sec.~\ref{sec:level4}. Furthermore, it can be used to estimate the approximate aperture in a configuration where a detailed optics does not yet exist, assuming that the $\beta$-functions at the bottleneck scale with $1/\beta^*$. This model has been used in the LHC to estimate the \bs reach for fixed beam-beam separation configurations and using achieved emittance values, allowing the performance of the machine to be pushed in order to deliver higher peak luminosity for the physics experiments.

\section{\label{sec:level3}Beam-based aperture measurement techniques}
The goal of the aperture measurements is to identify and quantify the real aperture restrictions in the machine for both beams and both transverse planes, or to quantify the aperture at given locations (usually for sensitive elements). The results can be used to optimise, whenever possible, the available clearance of the beams. Aperture measurements are performed after having established the reference closed orbit and corrected the optics, and are used as input to define the settings of the collimation system that should guarantee protection of the measured aperture.

In this section, we describe all beam-based methods for aperture measurements using movable collimator jaws that have been developed during the operation of the LHC. These methods are generic and can be applied in other circular accelerators. The different techniques are based on the general principle of generating losses with a safe beam, i.e. beam conditions that are deemed non-dangerous for the ring integrity. These losses are measured by means of the beam loss monitors (BLMs)~\cite{holzer05,holzer08a} placed all around the ring. Different analysis and reconstruction methods are performed, depending on the method and type of aperture measurement (global or local), in order to derive the aperture at the bottleneck location from these losses in units of \sig. 

\subsection{Emittance blow-up with resonance crossing and collimator scans}

In this method, the transverse beam emittance is blown-up in a given plane by crossing the corresponding 3$^\text{rd}$ order resonance, causing beam losses that are detected by the BLMs around the ring. This method combines the beam excitation technique with collimator scans (CS), in which the aperture is scanned with the movable collimator's jaws. 

\begin{figure}[!htbp]
   \centering
   \includegraphics[width=10cm]{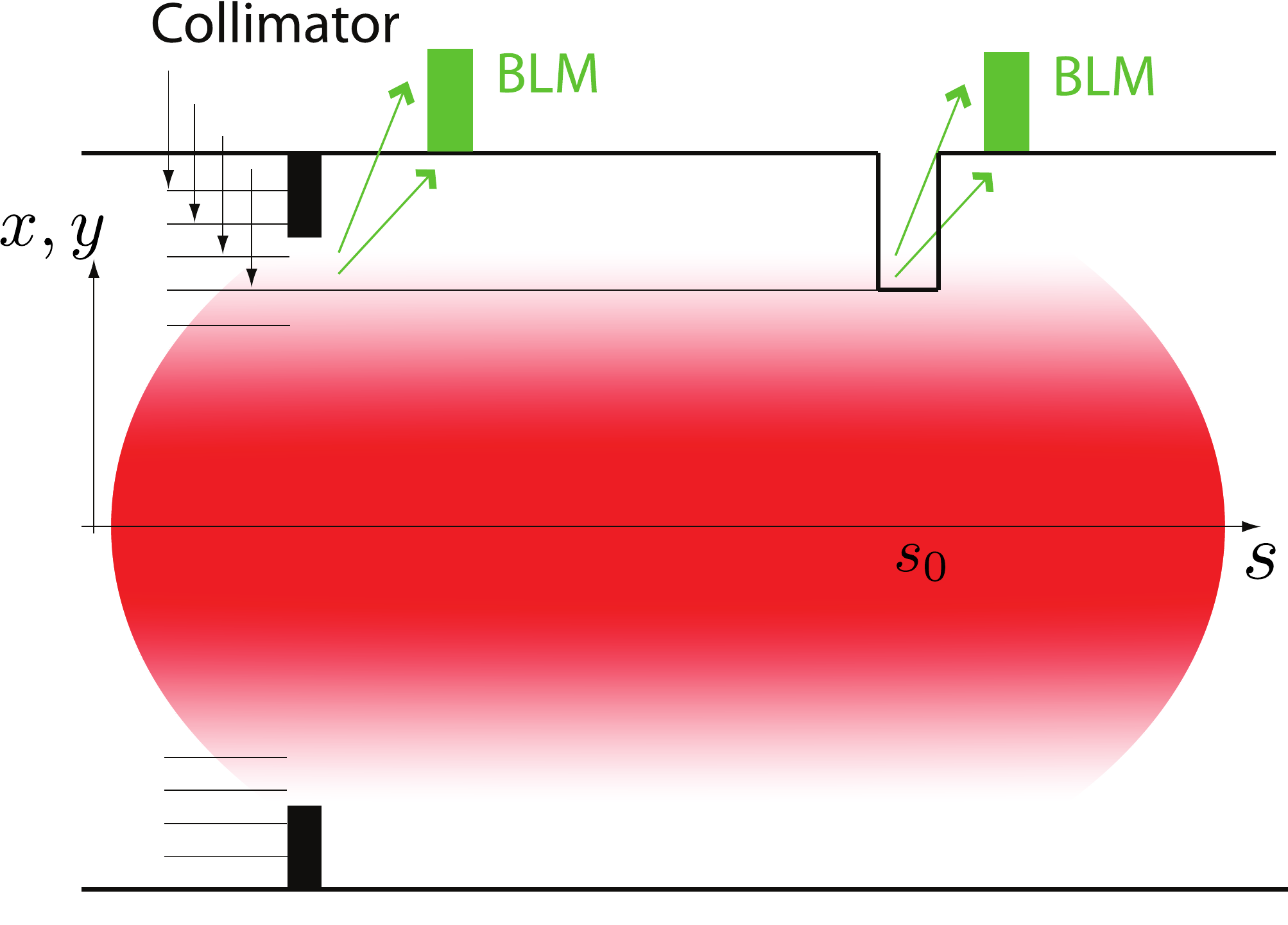}
   \caption{Illustration of the beam-based aperture-measurement technique combining beam blow-up and collimator scans (the reference collimator is shown in black and the BLMs in green).}
   \label{fig_SchemeBU}
\end{figure}

The initial step is performed with all collimators opened enough to expose the global bottleneck, the reason why this technique requires beams of very low intensity. In these conditions, the beam emittance is blown-up by approaching the tune to the third-order resonance using trim quadrupoles, until beam losses are observed around the ring, so identifying the global aperture bottleneck. The beam becomes so degraded after the tune change that it is not usable anymore, and a fresh beam has to be injected. The blow-up is repeated while a reference collimator, typically a primary or a tertiary collimator acting in the plane to be probed, is closed in steps of 0.5~\sig until the aperture bottleneck is shadowed by the collimator with beam losses recorded at the BLM closest to the reference collimator, as illustrated in Fig.~\ref{fig_SchemeBU}. The collimator used as reference has to be centred around the reference orbit before starting the scan. The collimator opening at which the global loss location moves from the ring to the collimator is used as an estimate of the magnitude of the global aperture bottleneck.
This CS provides the beam loss rates $R_\mathrm{loss}^\mathrm{ring}$ at the aperture bottleneck in the ring and at the reference collimator, $R_\mathrm{loss}^\mathrm{coll}$ for different values of the collimator half-gap $A$. Because of unavoidable intensity variations during the measurements, the raw BLM data are normalised by the number of protons lost measured by means of the beam current transformer (BCT). In addition, different BLM responses due to different optical functions, different positioning of the BLMs, and different geometries and materials can cause significant variations between the BLM signals recorded at different locations in the ring. Because of this, each BLM signal is normalised by the maximum value obtained when the aperture bottleneck is fully exposed. 

\begin{figure}[!htbp]
   \centering
   \includegraphics[width=10cm]{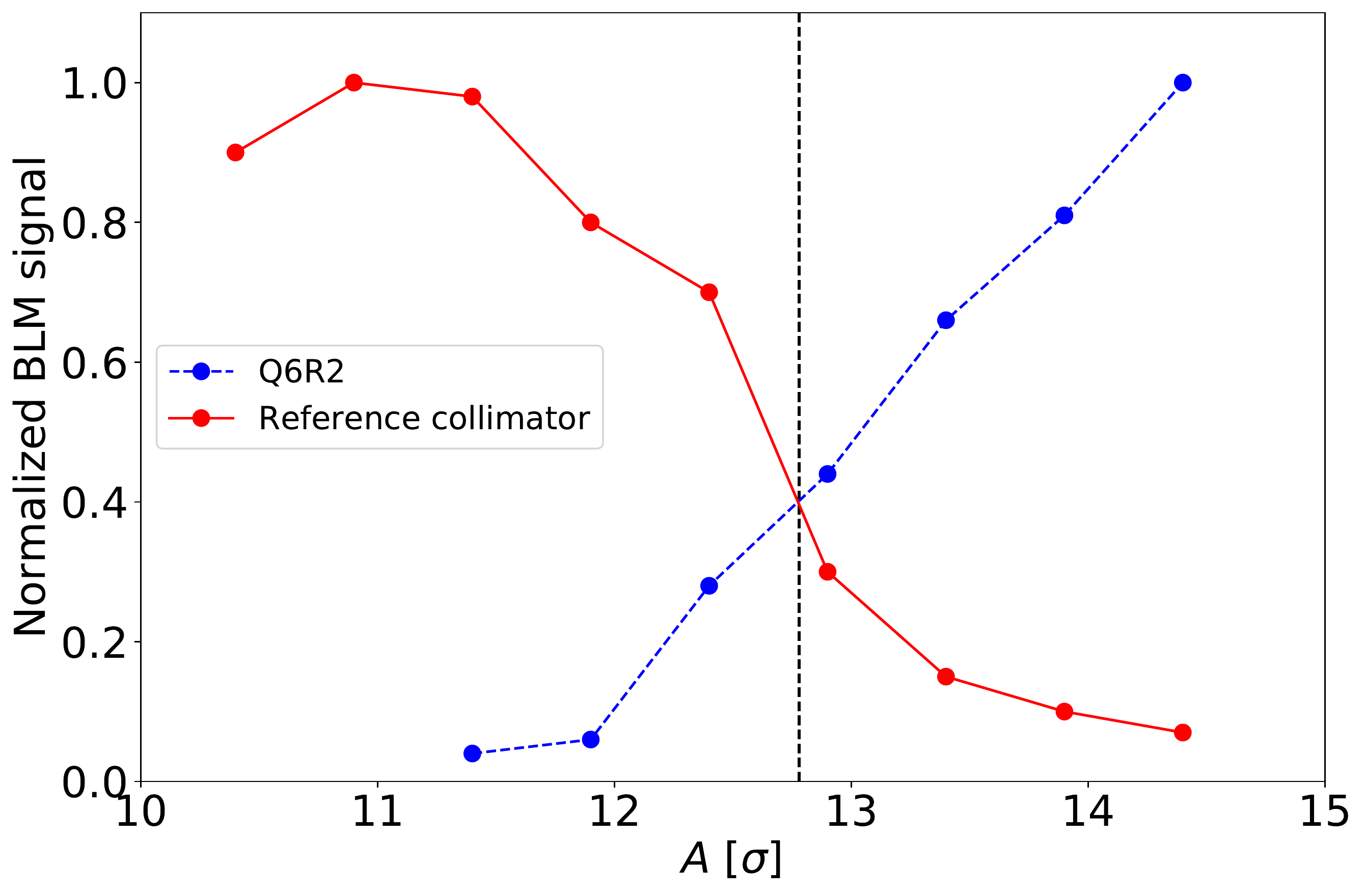}
   \caption{Normalised BLM signal measured at the reference collimator (red curve) and at the aperture bottleneck (blue curve) as a function of the half-gap of the reference collimator. Equal aperture at the bottleneck and at the reference collimator is indicated with a dashed line. This example corresponds to a measurement performed in 2017 for B1 in the horizontal plane with the injection optics (450~GeV).}
   \label{fig_Scheme_coll_scan}
\end{figure}

An example of the resulting normalised loss rate, $\tilde{R}_\mathrm{loss}$, as a function of the reference collimator half-gap is plotted in Fig.~\ref{fig_Scheme_coll_scan}. Equal signals at the quadrupole Q6R2, in this case identified as the global bottleneck, and at the reference collimator are indicated with a dashed line. This is used to infer the interpolated aperture value of the bottleneck in units of \sig. It should be noted that all imperfections are naturally taken into account and a fully realistic on-momentum aperture is obtained.

These measurements can be performed on-energy as well as for any specified off-momentum configuration. This method is very powerful, although limited to injection conditions, as new beam needs to be injected after each tune-change manipulation. Measurements at top-energy would require an energy ramp after each re-injection of the beam, which becomes prohibitive in terms of time. Moreover, the tune change is a global manipulation that affects each bunch in the beam. Therefore this approach prevents the selective use of individual bunches.

The accuracy of this measurement technique is given mainly by the collimator step size, which is typically chosen to be 0.5~\sig, and by the accuracy on the centring of the reference collimator around the reference orbit. Another possible systematic uncertainty comes from the potential ambiguity in determining the beam loss level where the losses due to the reference collimator opening and those at the bottleneck are equal. When these signals are at the same level, the bottleneck and reference collimator are intercepting both the beam and the secondary halo. By moving the reference collimator jaws a small step inwards, one can verify if the collimator is really touching the beam edge. Operational experience in the LHC shows a distinct step change of the BLM signals when the beam is intercepted, which makes this uncertainty small compared to the step size of the jaw movement used in the aperture measurements.

Note also that any $\beta$-beating at the collimator can also affect this measurement technique. If the nominal $\beta$-function at the collimator is used to calculate its opening in \sig, any deviation of the local optics for the theoretical one would alter the real beam size and hence the estimated aperture. To avoid this, the collimator opening should be calculated using the measured $\beta$-function value from previous optics measurements, which nevertheless also has an associated measurement error. However, what usually matters for machine protection considerations is the relative opening of the aperture bottleneck and of the collimator. This quantity is directly inferred from the measurements without the need of further corrections. 

This method was used in the LHC from 2010 to 2012 for global aperture measurements at injection energy (450~GeV)~\cite{giovannozzi10lhccwg,ipac11_assmann_aperture} and is applicable to any other circular accelerator equipped with collimators with movable jaws.

\subsection{Emittance blow-up with controlled white noise and collimator scan}
This technique was developed with the main goal of better controlling the beam losses, so that the same beam can be used for several subsequent excitations, making it possible to perform aperture measurements at top-energy in an efficient manner. It combines a controlled emittance blow-up, achieved by means of a white-noise excitation generated by the LHC transverse damper, with the CS used in the previously described method.

As the transverse damper excitation can be gated on individual or groups of bunches with a sufficient longitudinal spacing, it is possible to selectively blow-up the emittance of individual bunches. This allows this method to be efficiently used at top-energy, as each bunch in a train of bunches can be used to perform a different measurement, eliminating the need to dump and re-inject the whole beam each time a new measurement is made. Primary collimators are typically used as reference collimators for global aperture measurements, while tertiary collimators are used for local aperture measurements at the triplet quadrupoles. This method was developed in 2012, to perform global aperture measurements in the LHC at top-energy. Note that this method is applicable to other machines, but it depends on the capabilities of the BLM system, as for the previous method, and of the transverse damper, the performance of which has a quantitative connotation.

A further improvement with this method is that the CS can be performed in reverse, i.e. starting with the reference collimator jaws in the innermost position with beam losses concentrated at its location. From an operational point of view this is a much safer approach, since the ring aperture bottleneck is only exposed to losses in the very last step, and hence for a shorter time period, thus minimising the potential risk of damage. 

The ensuing analysis of the data collected and aperture interpolation is performed following the same procedure as for the previous method, illustrated in Fig.~\ref{fig_Scheme_coll_scan}.

\begin{figure}[htb]
   \centering
   \includegraphics*[width=9cm]{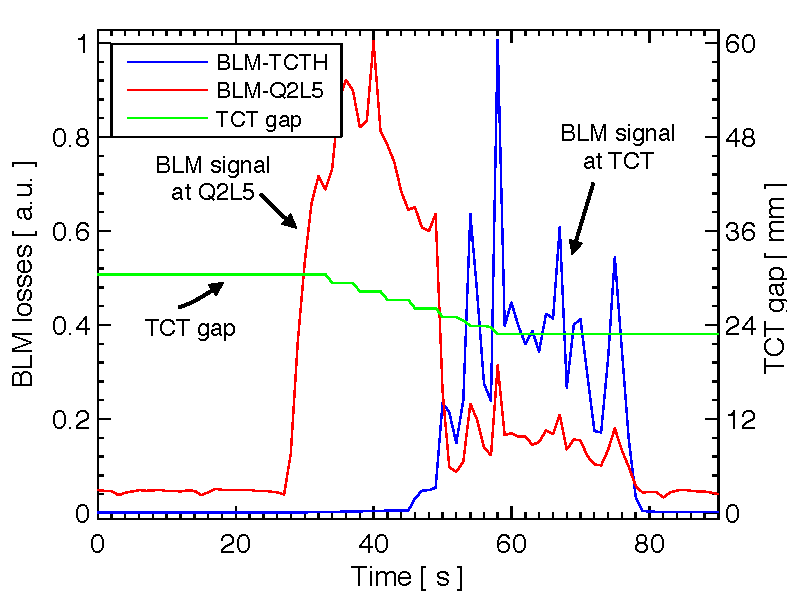}
   \caption{Losses at the reference TCT and at the triplet magnet Q2L5 and TCT full gap in mm during a semi-automatic aperture measurement 
     at 4~TeV, with $\beta^*\,=\,60$~cm. The transverse damper excitation lasted from 26~s to 80~s. }
   \label{figADTtopEn}
\end{figure}

The time required to perform these measurements was later reduced by automating the procedure. The transverse damper can be set up to excite the beam continuously but mildly over an extended period of time, while the reference collimator jaws can be programmed to open automatically in steps of 0.5~\sig at a defined frequency. Once the aperture bottleneck is exposed, the measurement is stopped. This is known as a semi-automatic aperture measurement. The application of the method at top-energy using the TCTs as reference collimators is illustrated in Fig.~\ref{figADTtopEn}. This shows a measurement performed with B1 in the horizontal plane. The BLM signals at the magnet Q2L5 (red) and at the horizontal TCT (blue) as well as the TCT full gap in mm (green) are depicted versus time. The TCT full gap of about 26~mm for which the losses at the TCT exceed the losses at the triplet corresponds to an aperture bottleneck of about 11.5~\sig.

In practise, this method can also be applied to confirm the measurement results from a manual CS, which is first used to identify the global aperture bottleneck in a controlled and safe way. Note that the frequency of the collimator jaw movement has to be chosen low enough to allow for the recording of the BLM signal at each step, and that a well-chosen combination of bunch intensity and transverse damper excitation is needed to produce losses large enough to provide a sufficient signal during the whole measurement procedure.

In the same way as for the emittance growth by means of tune change, the measurement accuracy is given again mainly by the step of the CS. Other contributions to the uncertainty are orbit shifts during the measurement (that can be monitored) and linear coupling between the two transverse planes. Thus, all high-order components must be minimised, meaning that the filling scheme must be chosen to avoid collisions and all higher order magnets from octupoles upwards must be switched off, with only sextupoles kept on in order to control chromaticity. Note that these sources of uncertainty affect all the beam-based aperture methods described in this paper.

\subsection{Emittance blow-up with controlled white noise and beam-based alignment}
The emittance blow-up with controlled white noise technique can also be combined with a beam-based alignment (BBA)~\cite{valentino12} instead of the CS in order to further speed-up the measurements. The BBA method was used in the LHC for the first time in 2015~\cite{hermes16_ipac_aperture}. In this approach all collimator jaws are retracted beyond the expected aperture bottleneck and the beam is excited with the transverse damper such that the losses are observed at the global aperture bottleneck. Once this is the case, it is the aperture bottleneck that then defines the beam envelope, which can be measured by performing a beam-based alignment with the reference collimator. In this type of alignment, the collimator is closed automatically and the movement is stopped by the alignment application~\cite{valentino12} when the signal of the BLM at the collimator exceeds a pre-defined threshold value. Once both collimator jaws have touched the beam, the normalised collimator gap can be directly translated into the available aperture at the bottleneck without any further post-processing. This method, known as a BBA measurement, is particularly suited for measurements in triplet magnets, where the settings of the local TCT required for protection can be directly derived from the measurement, with a resolution down to $0.1~\sigma$. 

As far as the possible sources of uncertainties are concerned, the beam edge may not be sharply defined and the measurement results might depend on the pre-defined threshold value of the BLM signal. This risks an overshoot of the collimator jaw position, meaning that the collimator jaw moves further in than the actual beam edge. As a consequence, this would cause an underestimate of the available aperture. There is also a risk that the collimator stops prematurely due to noise on the BLMs, but this can be easily managed by the control software. Nevertheless the result should ideally be validated by a reduced CS. This implies performing an additional emittance blow-up at the final collimator position, verifying that the losses are at the collimator, followed by a second blow-up after a small step outwards of the collimator jaw, e.g. by 0.1-0.5~\sig, to verify that the losses move back to the ring bottleneck.

\subsection{Local orbit bumps}
\begin{figure}[!htb]
   \centering
   \includegraphics*[width=9cm]{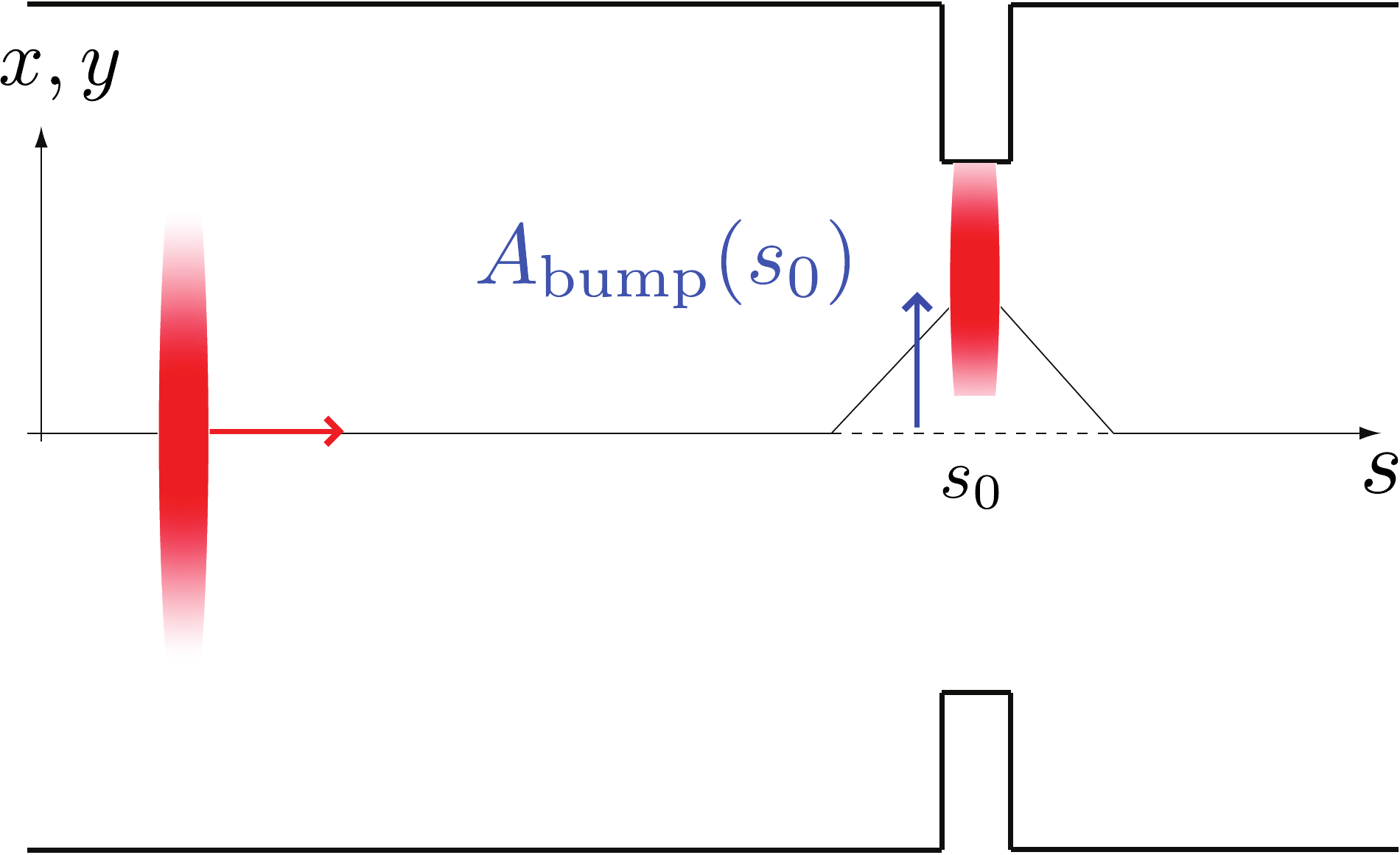}
   \caption{Sketch of the local closed-orbit bump method.}
   \label{fig_Scheme3C}
\end{figure}

In order to perform local aperture measurements at a  location other than that of the global bottleneck or to optimise the available aperture by centring the closed orbit at an identified global bottleneck, closed-orbit bumps are used. The closed-orbit bump is designed to have the maximum amplitude as close as possible to the location under study, as depicted in Fig.~\ref{fig_Scheme3C}. This is a classical approach to aperture measurements at the LHC, which is combined with CS.

The bump amplitude is slowly increased in both the positive and negative directions, until the beam touches the aperture and local losses are observed. This ensures a check in case of asymmetric apertures. Different bump shapes can be used, but typically a 3- or 4-corrector bump is sufficient to identify local restrictions. The distance $A_\mathrm{loc}$ in mm between the nominal beam centre and the aperture is the sum of the local bump amplitude $A_\mathrm{bump}$ and of the beam envelope $N\sigma$, with \sig evaluated at the location of the bottleneck:
\begin{equation}
 A_\mathrm{loc}=N\sigma+A_\mathrm{bump} \, .
 \label{eq:ap_bump}
\end{equation}

The amplitude $A_\mathrm{bump}$ can be obtained from the expected closed-orbit value for the settings of the bump, or from the measured orbit with the beam position monitors (BPMs). Of course, in case the BPM is not close enough to the location considered by the aperture measurement, $A_\mathrm{bump}$ should be reconstructed by using measured values at the BPMs and the transfer matrices from the BPMs to the location of the aperture measurement. In order to reconstruct the normalised aperture in units of the local beam size, Eq.~(\ref{eq:ap_bump}) can be divided by \sig.

As a prerequisite for these measurements, the normalised beam envelope $N$ should be defined. To do that, a primary collimator gap in the relevant plane is set at a pre-defined normalised value $N$, typically in the range 3--5~\sig. At this setting the beam is excited with the transverse damper until losses are observed. In this way the beam fills up the full space between the collimator jaws, up to $N\sigma$, with \sig taken at the primary collimator. The edge of this envelope will then touch the aperture at a normalised amplitude $N$ as the orbit bump is increased. Having a sharp edge to the beam distribution makes it easier to detect the losses when the beam touches the aperture. 

\begin{figure}[!htb]
   \centering
   \includegraphics*[width=10cm]{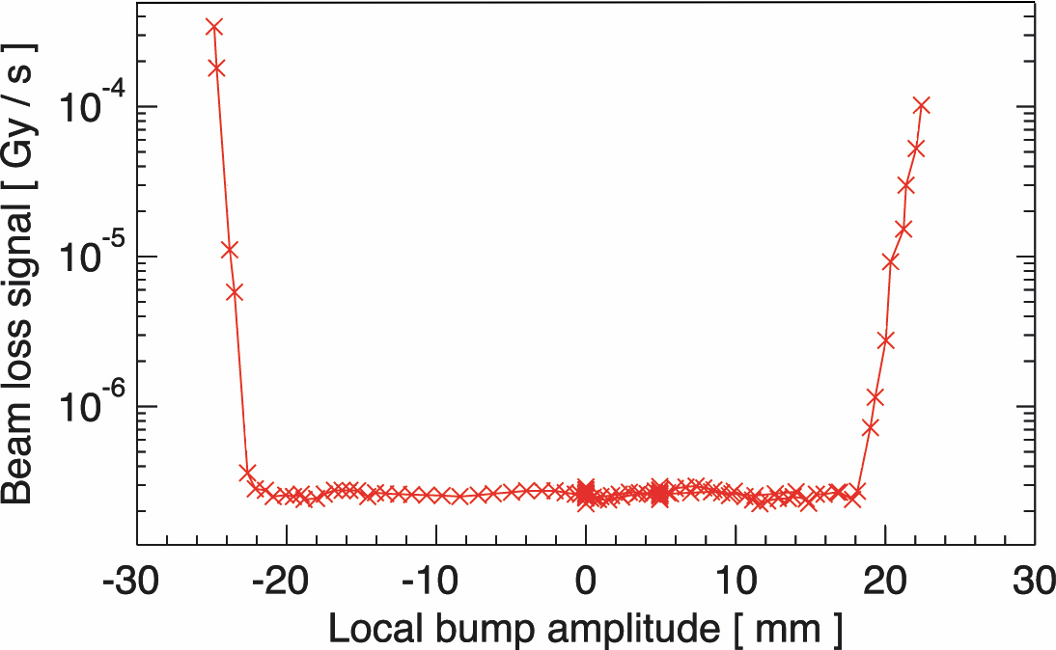}
   \caption{Example of measured beam losses at the quadrupole Q4L6 for B1 versus the orbit bump amplitude (performed in 2010). Note that the beam losses for the two signs of the bump amplitude depend on the amount of beam that is cut during the preceding measurements.}
   \label{fig_bump_blm}
\end{figure}

An example of the beam losses measured at the quadrupole Q4L6 as a function of the local orbit-bump amplitude for B1 is shown in Fig.~\ref{fig_bump_blm}. Note that the beam losses for the two signs of the bump amplitude depend on the amount of beam that is cut during the preceding measurements. Furthermore, in between each measurement, the normalised envelope $N$ has to be redefined using the TCP opening and the transverse damper excitation. 

An intrinsic advantage of this technique is that possible asymmetries in the aperture shape can be identified by performing measurements with both signs of the bump. This then gives the possibility to deploy a permanent orbit bump to centre the beam at the bottleneck location and hence, gain aperture for machine operation. 

\begin{figure}[!htbp]
  \centering
  \includegraphics[width=8cm]{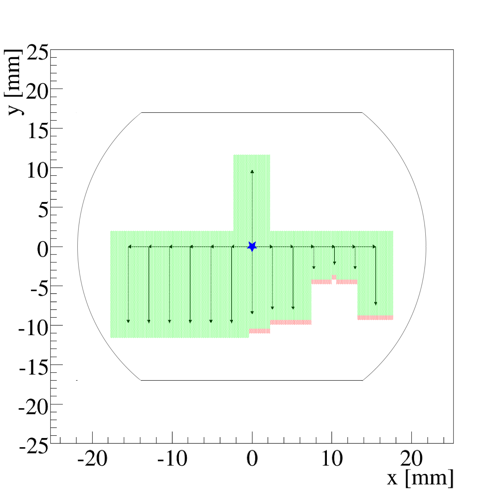}
  \includegraphics[width=6cm]{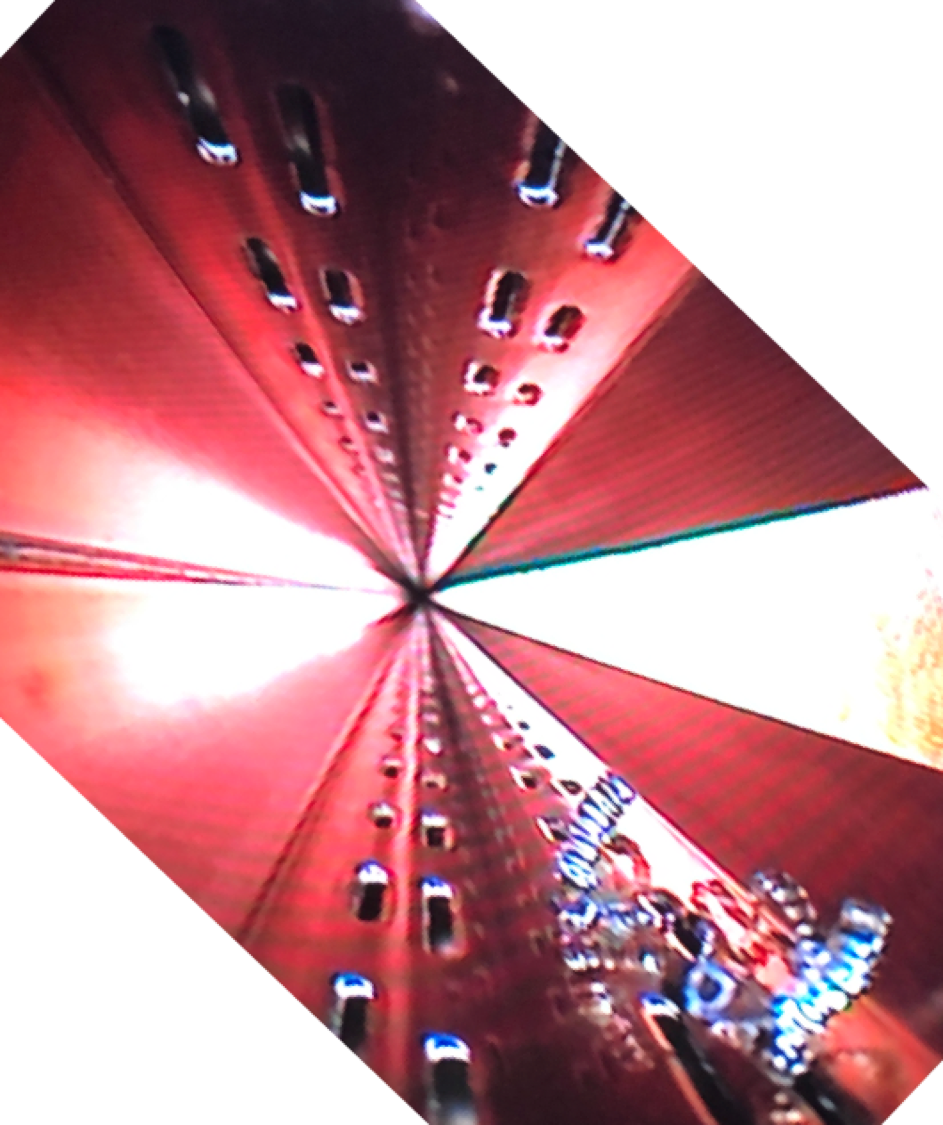}
  \caption{Left: Outcome of the 2018 ULO aperture scan. The beam screen is represented by the black line, the black arrows indicate the displacement of the beam centre performed using local orbit bumps, the reference orbit at the moment of the measurement is shown by the blue star, the clear aperture is reported in green while the measured edge of the ULO is shown by the red boxes with dimensions defined by the measurement resolution. The machine reference frame is used, with the B2 entering in the plane of the plots. Right: picture taken during the first endoscopy carried out in April 2019. The top and bottom flat parts of the beam screen with the pumping slots are clearly visible. The twisted material in the foreground of the lower-right part of the picture is the ULO.}
\label{fig:ULO}
\end{figure}

An example of this approach is given through an issue that arose during the Run~2 operation, with fast (typically of a few ms) beam losses happening in IR8 leading to 14 beam dumps, 3 of which caused a magnet quench. Energy-deposition studies showed that the vertex of the hadronic showers generating the limiting BLM signals were most likely situated within 1~m from the centre of the dipole MB.C15R8~\cite{Daniele:evian19}. In order to investigate the origin of these losses several local aperture measurements were performed in that area, which revealed the presence of an Unidentified Lying Object (ULO) (more details can be found in~\cite{ULOIPAC,Daniele:evian19}). A constant monitoring of the ULO position and orientation during Run~2 was carried out by means of local aperture measurements, which were performed at every beam commissioning at the beginning of each run. An example of the aperture scans performed in 2018 is shown in Fig.~\ref{fig:ULO} (left). The physical beam screen aperture is indicated by the black line. The black arrows follow the path of the beam centre from various positions in the horizontal plane using local orbit bumps, while the nominal reference orbit at the moment of the measurement is shown by the blue star. A clear aperture, i.e. no losses observed, is reported in green while the measured edge of the ULO is shown by red boxes with dimensions defined by the measurement resolution.

A mitigation strategy was adopted in the LHC based on the deployment of a local orbit bump that allowed the beam to stay sufficiently far away from the ULO, while ensuring at least 10~\sig of available aperture at injection energy. After the end of Run~2 an endoscopy revealed that the ULO was a twisted string of plastic a few cm in size, which was stuck between the pumping slots of the beam screen and the cold bore (see Fig.~\ref{fig:ULO}, right). The ULO was successfully removed.

In the LHC, this method was also used to investigate the triplet aperture in the experimental insertions  to conclude on the \bs reach. For this particular case, the method was combined with CS, using the TCTs as reference collimators in order to keep the triplet quadrupoles protected as much as possible and minimise their exposure to beam losses. With these collimators at their nominal settings additional crossing bumps were added on top of standard crossing and separation schemes until the beam touched the reference TCTs. The collimator jaws were then retracted and the bump amplitude increased in steps until the triplet aperture was exposed. This allowed the TCTs  settings that protect the triplet to be determined~\cite{Salvachua:2302435}. 

This method ensures triplet protection during the measurements as they remain in the shadow of the TCTs to about 0.5~\sig (TCT step size). On the other hand, the aperture results are expressed in terms of TCT retraction and have a small dependence on the detailed shape of the bump used to perform the scan, which is affected by the kicks generated by the transverse misalignment of quadrupoles or by the orbit correctors in between the TCTs and the triplets. A good initial alignment of the TCTs is mandatory in order to have accurate results.

Although rather robust and widely used, a potential risk with the orbit-bump technique is to generate large losses at the studied element if the steps of the orbit bump are too large. One way to mitigate this consists of using a variant where the orbit bump is used only to displace the global bottleneck to the local element that should be studied. In this way, one of the other techniques for a global aperture measurement can be used, e.g. the semi-automatic method or the BBA method, where the losses induced by blowing-up the emittance using the transverse damper are much better controlled. The aperture of the element under study is then given by the value obtained with the global aperture measurement minus the amplitude of the local bump introduced. This variant also has the advantage of typically using smaller orbit bumps, thus reducing the risk to run out of orbit corrector strength during the scan. 

Finally, it is worth mentioning that as a consequence of some of the first results obtained using the orbit-bump technique, it was decided to install additional BLMs in some IRs to increase the resolution of the method and investigate in more detail certain aperture restrictions~\cite{redaelli12_IR2aperture}. This stresses the importance of having a large quantity of BLM units around the accelerator to accurately determine the location of aperture bottlenecks.

\section{\label{sec:level4}Measurement results}
The detailed knowledge of LHC aperture bottlenecks achieved thanks to the improved accuracy of the aperture measurements has been a key ingredient for the LHC performance in terms of both peak and integrated luminosity~\cite{bruce17_NIM_beta40cm}. In 2018 the LHC was operated with \bs= 25~cm in the high-luminosity experiments, representing a huge gain compared to the nominal LHC operational scenario with \bs=55~cm. Together with the improvement of the injected-beam brightness, the dramatic reduction of \bs has been a major contribution to the excellent luminosity performance at the end of Run~2, achieving a factor 2 higher peak luminosity than the LHC design. 

\subsection{Aperture measurements at injection energy (450 GeV)}
Since 2010, global aperture measurements have been performed at the 450~GeV injection energy using the various methods described in Section~\ref{sec:level3}. Although not directly relevant for the LHC \bs-reach, the measurements were important to ensure aperture protection by the collimation system, since the aperture is more critical at injection for most elements due to the larger beam size, with the exception of the elements affected by the \bs-squeeze at top-energy. 

\begin{figure}[!htbp]
  \begin{centering}
  \includegraphics[width=14cm]{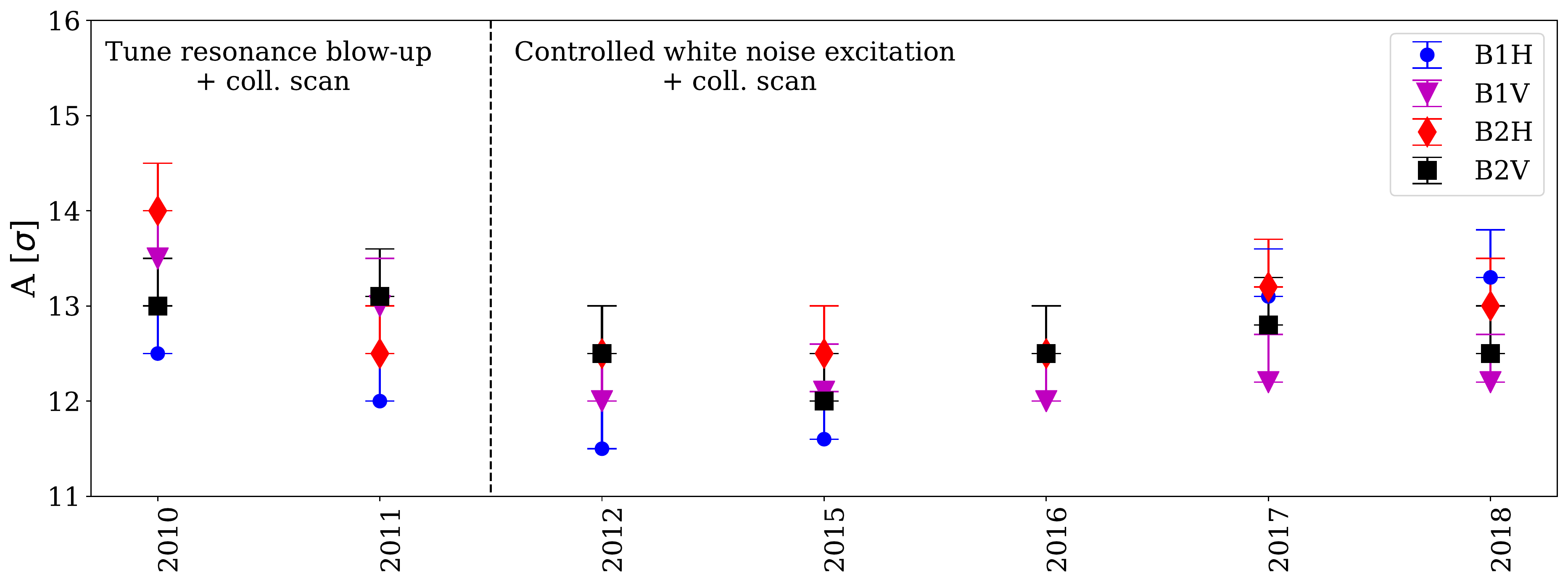}\\
  \end{centering}
  \caption{The measured global-aperture bottlenecks at injection (per beam and plane) during Run~1 and Run~2. All values are given in units of beam \sig, for \en=3.5~$\mu$m. }
\label{fig:injection_aperture}
\end{figure}

Fig.~\ref{fig:injection_aperture} shows a summary of the measured global-aperture bottlenecks in the two beams and planes at injection during Run~1 and Run~2. The aperture values depicted correspond to the maximum collimator jaw position opening in the scan before exposing the ring aperture. The associated error bars are determined by the step used during the collimator scan. Note that in 2010--2011 the tune-resonance emittance blow-up technique was used while since 2012 this has been performed using the controlled white noise excitation method. The observed differences over the years are probably due to a combination of effects such as variations in closed orbit and changes in the optics correction. The re-alignment of machine elements~\cite{bruce16_inj_aperture_HL} could also explain some of the observed differences, however there is not enough data from the survey measurements to pinpoint the possible sources. More details about the location of the bottlenecks and methods used to perform the measurements are given in Appendix~\ref{table_injection}.

In 2015, a comparative study was performed in order to verify the compatibility between the different methods developed for aperture measurements at injection. In this test the methods compared were: the transverse damper blow-up with CS; the transverse damper blow-up with BBA; the local orbit bump with CS. From these measurements we could conclude that all methods confirmed the same aperture bottleneck locations. A summary of the results obtained for both beams and both transverse planes is shown in Table.~\ref{injection_comp}. Note that the aperture values measured using the collimator-scan method are given by a range defined by the collimator jaw position in the scan before and after exposing the ring aperture. This applies for the CS and the local orbit-bump methods. For the BBA method the main error comes from the pre-defined BLM threshold at which the jaw stops that could make the measured aperture smaller. Hence, the error depends on the exact distribution of beam halo close to the cut, which is not well known, and which could change in different measurements, as well as the applied BLM threshold. It is also possible that, due to noise on the BLMs, the collimator jaw stops prematurely, especially if a low BLM threshold is used. Due to the origin of these uncertainties, it is not easy to associate a meaningful number for the uncertainty on this method. These measurements should ideally be validated by a reduced CS combined with emittance blow-up, and the agreement with the other methods can be seen in Table~\ref{injection_comp} where a summary of all the results obtained is presented.

\begin{table}[!htbp]
  \begin{center}
    \caption{2015 measured values of global aperture bottleneck at injection energy, expressed in local beam \sig with the underlying assumption of a normalised reference emittance of $3.5~\mu$m.}
    \begin{tabular}{lcccccc}
      \hline
      \textbf{Beam}&\textbf{Plane}&\textbf{Bottleneck}&\textbf{BBA}& \textbf{CS}& \textbf{Local orbit bump} &\textbf{Local orbit bump} 
      \\
            &&& & &\textbf{Positive sign} &\textbf{Negative sign} 
      \\\hline\hline
      B1&H        & MBRC.4R8  &11.6 & 12.5-13.0 &NA &NA\\ 
      B1&V         & Q6L4 &12.1 & 11.5-12.0&12.5-13.0&14.5-15.0  \\ 
      B2&H         & Q4L6  &12.5& 12.8-13.3&13.0-13.5 &13.5-14.0 \\ 
      B2&V         & Q4L6  &12.0& NA &12.5-13.0 & 12.5-13.0\\ \hline
    \end{tabular}
    \label{injection_comp}
  \end{center}
\end{table}

Note that for B1 in the horizontal plane the global aperture bottleneck was found in a separation dipole in IR8, however at this location it was not possible to create an appropriate orbit bump. Hence, the measurements could only be performed with the transverse damper blow-up techniques. Moreover, for B2 in the vertical plane, the CS data was corrupted, so the aperture was evaluated only with local measurements and the BBA method. 

In conclusion, all methods agree well on the bottleneck location. The CS and local orbit bump methods agree within their estimated associated error of 0.5~\sig. Differences up to 0.9~\sig are observed between the measured aperture values in units of~\sig between the BBA method and the other two methods, which could originate from the pre-defined BLM threshold and BLM signal noise.

\subsection{Aperture measurements for physics configuration}
Before global aperture measurements were operational at top-energy towards the end of Run~1, the local orbit-bump method was used in the LHC to investigate the \bs reach in the high-luminosity insertions (IR1 and IR5). The limiting aperture bottleneck is expected to be in the triplets of IR1 and IR5 with squeezed beams for \bs smaller than 5~m, or slightly higher if the machine imperfections are unfavourable. The first measurements performed in 2011 and 2012 allowed a change of \bstar from 1.5~m to 1~m in IR1/5~\cite{redaelli11evian}, which could be further reduced to 0.6~m in 2012 thanks to tighter collimator settings~\cite{bruce15_PRSTAB_betaStar}. The beam-based aperture measurement technique exploiting controlled white noise emittance blow-up, was developed in 2012 and since then it has become the standard method for aperture measurements at the LHC. 

A detailed summary of all aperture measurements performed for physics configuration with squeezed beams is shown in Table~\ref{tab:aperture_collision_local} and Table~\ref{tab:aperture_collision} in Appendix~\ref{table}. Both the aperture bottleneck location and its value in units of \sig are compiled for both beams and both planes (for reference the optics configuration and beam energy are also listed). Note that the aperture values are again given by a range defined by the collimator position in the scan, before and after exposing the aperture. As can be seen in Table~\ref{tab:aperture_collision} the bottlenecks are always found in the triplet regions of IR1 or IR5 as expected. In most cases the same location remains the bottleneck for a given beam and plane. It should be noted that when the bottleneck is indicated to be in Q3/D1 the highest loss in the measurement is found at the BLM labelled as Q3 but is actually installed close to the interconnection between Q3 and D1 and is likely to be more affected by showers from the upstream D1 than from losses within Q3 for the incoming beam. From theoretical aperture studies, the bottleneck is also much more likely to be in the D1.

\begin{figure*}[!htbp]
   \centering
   \includegraphics*[width=15cm]{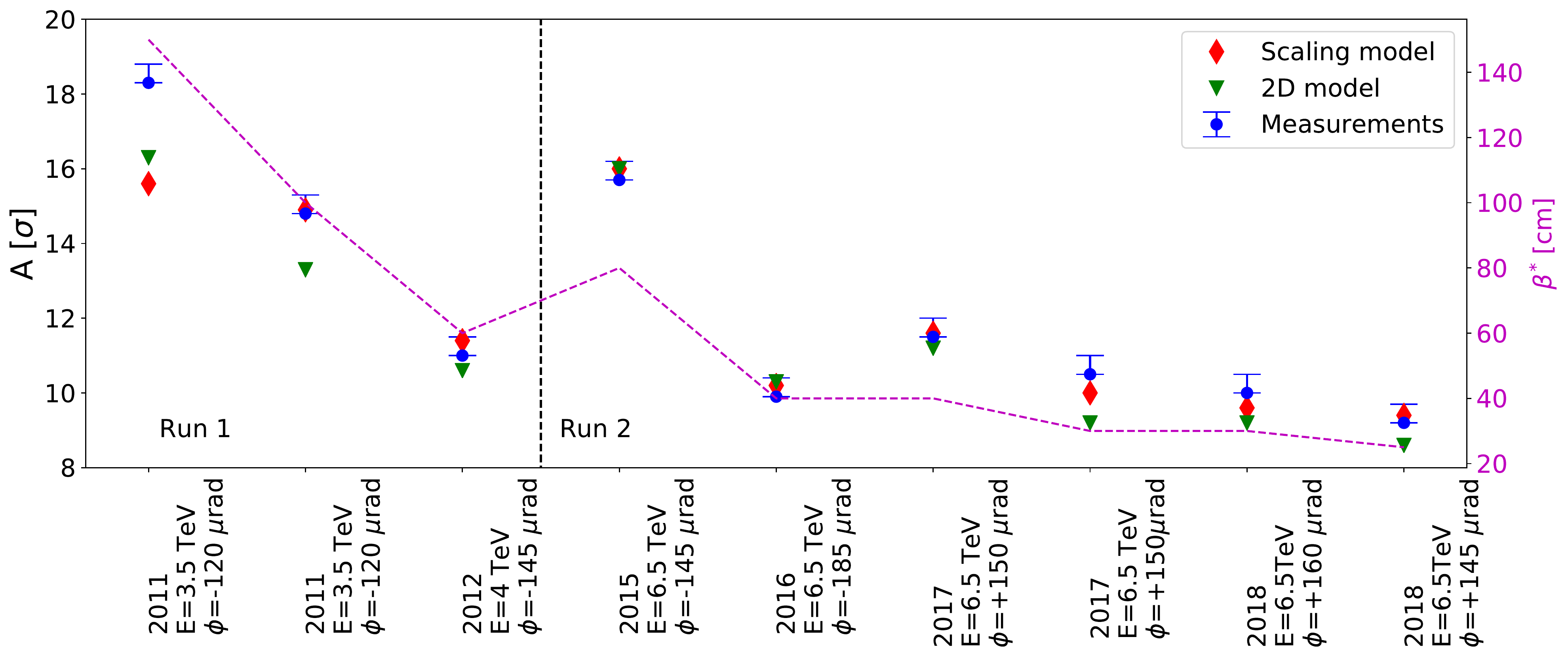}
   \caption{Top-energy Run~1 and Run~2 aperture values calculated with the $n_1$ and the scaling methods together with the measured values (only the minimum over the two beams and planes shown). The energy, $E$, and crossing angle, $\phi$, are indicated.}
   \label{summary_global_aperture}
\end{figure*}

Fig.~\ref{summary_global_aperture} shows the Run~1 and Run~2 aperture bottlenecks for the physics configuration from the $n_1$ and the scaling models, together with the measured values (only the minimum over the two beams and planes is depicted for each optics configuration). Note that for the first aperture measurements at top-energy, performed in 2011, the scaling method takes into account the smallest aperture measured at injection energy, which is not at the triplet quadrupoles as their aperture is larger than the global injection bottleneck. Therefore the 15.6~\sig is a pessimistic value. The aperture values presented in Fig.~\ref{summary_global_aperture} correspond to the collimator aperture before exposing the aperture of the bottleneck. The real aperture will be between these values and these values plus 0.5~\sig, corresponding to the step of the scan. Because of this a 0.5~\sig error has been associated to the aperture values in Fig.~\ref{summary_global_aperture} only on the upper side of the values. 

As far as the $n_1$ model is concerned, during the LHC Run~1, several potential sources of error were found to be less serious than the design assumptions as explained in Section~\ref{sec:level2}, and the tolerances for the $n_1$ calculations had to be adjusted as summarised in Table~\ref{n1_param}. The new tolerances based on measurements were used for aperture calculations since 2012, explaining the different agreement between the $n_1$ method and the measurement results when comparing the first two points (about 2~\sig) and the others (within 0.5-1~\sig). Apart from these initial points we observe that the predicted aperture using the scaling model is never further away than 0.5~\sig from the measured aperture. These results are consistent with the fact that the step size in the CS was 0.5~\sig for most measurements. We therefore take this as a guide to the uncertainty of the method. We consider this a very good agreement given the collimator step size and the unavoidable small differences in orbit and optics corrections between years.
 
\begin{figure}[!htbp]
  \begin{centering}
  \includegraphics[width=10cm]{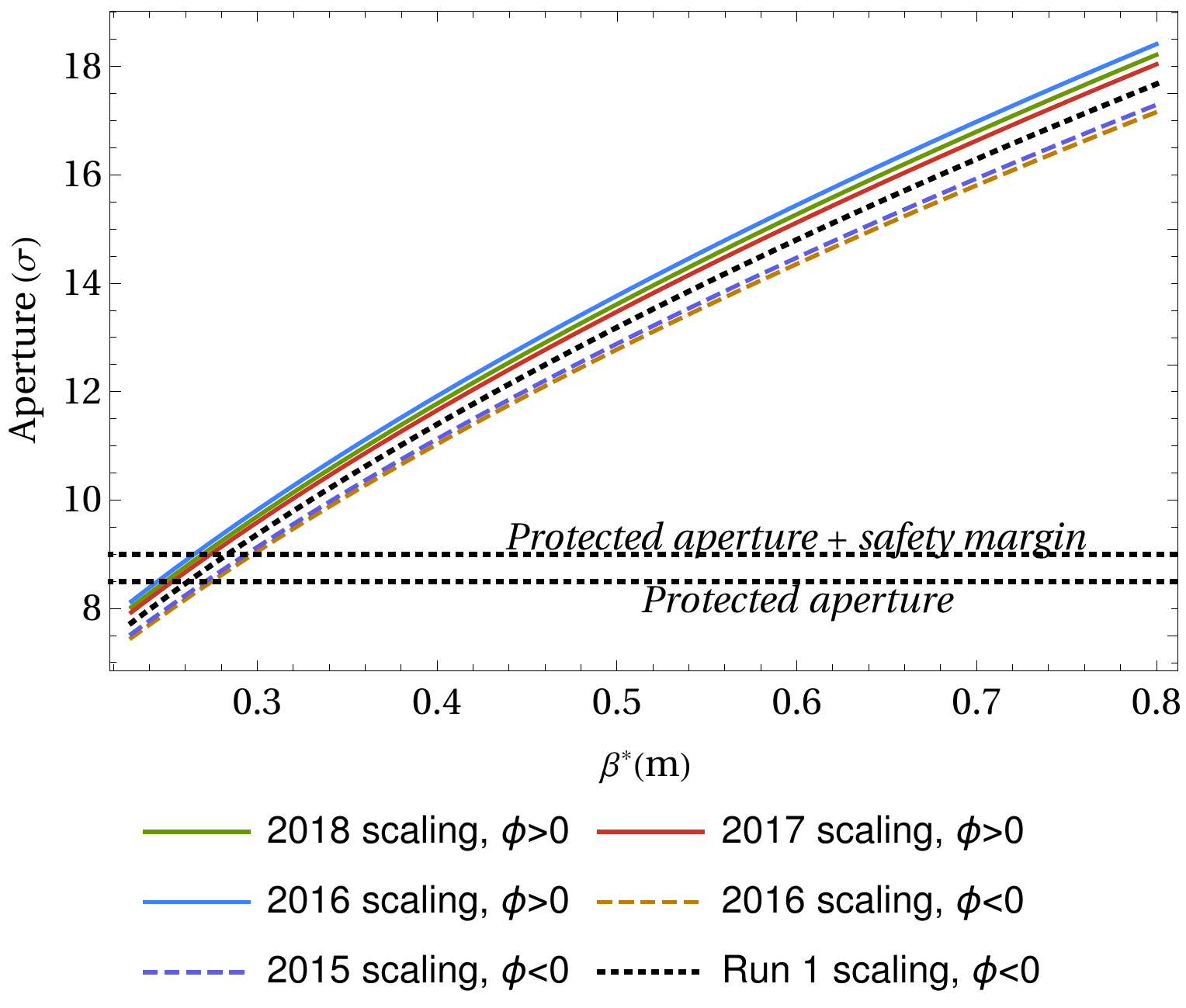}\\
  \end{centering}
  \caption{Triplet aperture as a function of \bs, calculated with the aperture-scaling method and using measured apertures from different years as input. These measurements were performed with positive and negative IR1 crossing angle $\phi$. All values are given in units of beam \sig, for \en=3.5~$\mu$m. }
\label{fig:collision_aperture}
\end{figure}

The aperture measurements reported in Fig.~\ref{summary_global_aperture} are not trivial to compare directly, since they were carried out under different machine configurations. In Fig.~\ref{fig:collision_aperture}, the estimated aperture as a function of \bs is shown. Each curve is calculated using the aperture-scaling method, but taking as input a different aperture measurement. All years in Run~2 have been considered, as well as the Run~1 measurement from 2012. For all curves, the beam-beam separation in units of beam \sig is assumed constant at the 2018 value (10.6~\sig for a normalised emittance \en=1.9~$\mu$m), with the crossing angle varying along the curves. 

All curves based on Run~2 measurements lie within 0.5~\sig with respect to their average no matter the sign of the IR1 crossing angle, $\phi$, which confirms the estimated uncertainty of the method. Based on this result, a safety margin of 0.5~\sig has been added on top of the calculated aperture starting from 2017 to ensure that the aperture declared for a new machine configuration is safe for regular beam operation. This approach minimises the risk of being forced to change the machine configuration after the beam commissioning has already started, which would cause a delay in the physics run due to the need to repeat some of the commissioning steps. Figure~\ref{fig:collision_aperture} shows a horizontal line representing the aperture that could be protected by the collimation system, using the 2018 collimator settings, and another line that includes the 0.5~\sig safety margin. A reach in \bs to a minimum of about 27.5~cm can be seen directly in the figure by comparing this line with the calculated aperture. In addition one can observe a systematic effect on the aperture as a function of the crossing angle polarity in IP1 (note that the crossing angle polarity in IP5 can not be changed). This is due to the observed increase of the vertical aperture measurements for positive crossing angles in IR1 with respect to the negative crossing angles. This can be explained by an asymmetric aperture model in the triplets region which could be due to mechanical errors or a misalignment of the magnets. 

\subsection{Reconstruction of physical apertures from physics configuration measurements}

From the measured aperture at a given location, $A_{x,y} (s)$, expressed in units of beam size, one can calculate the corresponding aperture in m, as

\begin{equation}
    A_{x,y}[\mathrm{m}](s)=A_{x,y}[\sigma](s) \times \sigma_{x,y}(s)+{\rm CO}_{x,y}(s) \, ,
    \label{eq:aper_mm}
\end{equation}
where $\sigma_{x,y}$ is the horizontal or vertical beam size (in m) and $CO_{x,y}$ the horizontal or vertical closed orbit (in m), at the location where the aperture has been measured. Both the $\beta$-function, which is used to compute the beam size, and closed orbit have been calculated using MAD-X for the corresponding optics configuration. The normalised nominal emittance of $3.5~\mu$m has also been used. Note that the triplet magnets are about $6$~m long and, due to the change of the $\beta$-function along their length, a variation of up to $4$~mm on the resulting aperture $A$ is expected between their entrance and the exit. The BLM system allows us to identify the element where the losses occur, but as only a few BLM monitors are attached to each magnet, it is not possible to accurately know the location of the loss inside the magnet. For the calculations presented in this section, the optics parameters in the middle of to the magnet location have been used.
Furthermore, we have computed the uncertainty related to the variation of the optics parameters over the magnet length as the difference between the extreme and mean aperture values inside the magnet.  

\begin{figure}[!htb]
   \centering
   \includegraphics*[width=14cm]{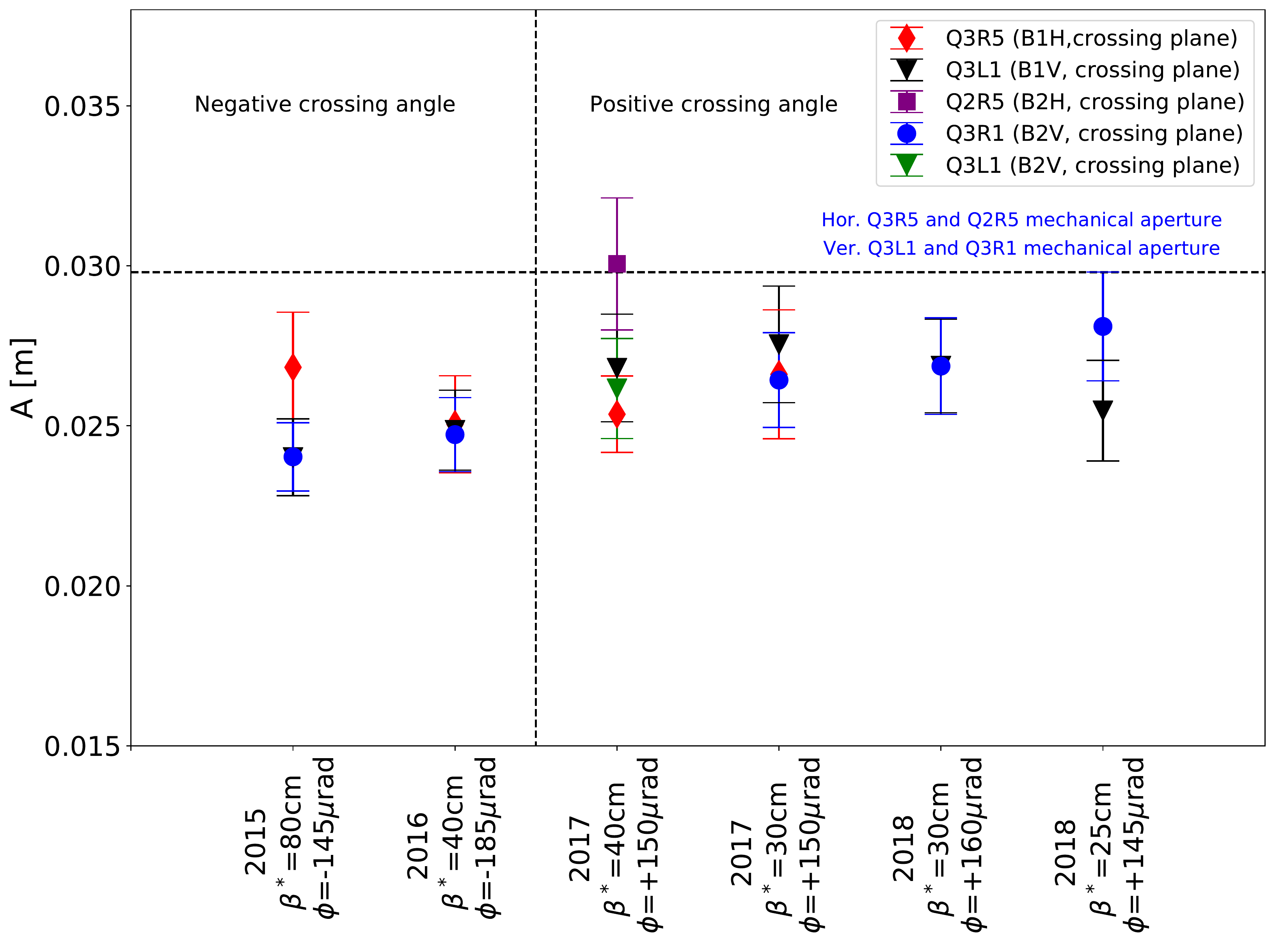}
   \caption{Crossing plane aperture measurements computed in m in the LHC for Run~2 in physics configuration. Different markers correspond to different elements and different colours correspond to different beams and planes. The horizontal dashed line indicate the mechanical apertures of the Q3R5, Q2R5, Q3L1 and Q3R1 magnets in the crossing plane. The vertical dashed line separates the measurements with negative and positive crossing angle in IP1.}
   \label{summary_global_aperture_mm}
\end{figure}

Fig.~\ref{summary_global_aperture_mm} shows the aperture calculations for all bottleneck locations found in the crossing plane as from Table~\ref{tab:aperture_collision}. The different markers correspond to different magnets, and the different colours correspond to the different beams and planes. The horizontal dashed line indicates the mechanical aperture of the Q3R5, Q2R5, Q3L1 and Q3R1 magnets in the crossing plane, while the vertical line separates the measurements performed with negative and positive crossing angle polarity in IP1. The error associated to each aperture value has been computed as the square root of the sum of the squares of the measurement errors given by the step of the scan, the uncertainty on the loss position and the measured $\beta$-beating. For the $\beta$-beating contribution the peak values have been used from Refs.~\cite{PhysRevAccelBeams.22.061004, PhysRevAccelBeams.20.061002,Private}. For some cases the $\beta$-beating peak value is not reported in these references, in such cases the $\beta$-beating peak has been calculated as three times the $\beta$-beating RMS value. Note that the aperture measurement value used for the calculation corresponds to the intermediate value between the collimator half-gap before and after exposing the bottleneck, which gives an uncertainty of 0.25~$\sigma$, corresponding to the reference collimator half-step used in the scan.

As can be seen in Fig.~\ref{summary_global_aperture_mm}, all calculated values are consistent with a slight loss in aperture compared to the physical aperture except one point that lies at the limit. A maximum aperture loss of about $6$~mm can be observed. Furthermore, a slight aperture gain linked with the change of polarity of the vertical crossing angle in IR1 can be observed, as was anticipated in Section~\ref{sec:level4}. The Q2R5 magnet measurement (square purple point) which is at the limit is made of two magnets named Q2R5.A and Q2R5.B, respectively (see Fig.~\ref{IPscheme}) and the beam losses were observed at $BLMQI.02R5.B2I21\_MQXB$, which is located in the middle of the Q2R5.A magnet. For this particular case, the variation of the $\beta$-function along the magnet means a change of aperture value from $0.031$~m at the entrance of the magnet (above the mechanical aperture) to 0.026~m at the end of the magnet (within the mechanical aperture). This indicates that the restriction should be located towards the middle of the magnet, which is consistent with the location of the BLM that measured the losses. 

For some measurements listed in Table \ref{tab:aperture_collision}, the aperture bottleneck was found at an unexpected location, e.g. at the IR1 triplet for a horizontal measurement, and at the IR5 triplet for a vertical measurement, i.e. in the separation plane. Due to the presence of the crossing angle in the vertical (IP1) and horizontal (IP5), it is expected that the horizontal bottlenecks are found in IR5, and the vertical ones in IR1. In addition the mechanical aperture in the non-crossing plane is significantly larger. For these measurements, the aperture in m was also computed using Eq.~(\ref{eq:aper_mm}), but the calculated values result in an unrealistically small aperture, i.e. about $15$~mm, which corresponds to about half of the mechanical aperture. One possible explanation to this observation could be the transverse nonlinear coupling present in the machine, which was unknown at the times of the measurements, and could turn a fraction of the excitation amplitude in one plane into an offset in the other plane, which eventually reaches the bottleneck in the non-excitation plane. The observed losses stopping the aperture measurement scan would then be in the non-excitation or skew plane, preventing the aperture in the plane being measured from being fully probed. Further investigations are needed to quantify the impact of the nonlinear coupling on the measurements and to better understand the cause of these observations. More work will be devoted to this aspect during the LHC Run~3.

\section{Conclusions}\label{sec:conclusions}
Aperture measurements in circular particle accelerators play an essential role for safe operation and to keep beam losses under control, the latter being a mandatory condition in superconducting rings. At circular colliders, like the LHC, reliable aperture measurements contribute to the boost of the performance in terms of peak luminosity. It should be noted that the small \bs, the high-intensity, and the high-energy of the beams used, which are all key parameters to increase the overall performance of the machine, define very powerful beams and even a tiny fraction of beam losses could cause a quench of the SC magnets or severe material damage. This makes the aperture measurements a genuine challenge, thus calling for an in-depth review and improvement of the standard methods used to probe the beam aperture. 

Various beam-based aperture measurement methods have been developed in the LHC over the years, with increasing precision and control over the beam losses generated. The methods presented in this paper are also directly applicable to other accelerators with similar constraints. Several of these are based on exciting the beam to blow-up the transverse emittance so as to generate losses which are then used to identify the bottleneck location, highlighting the importance of a versatile transverse damper system. The aperture measurements from all these methods are in general agreement have been shown to provide reproducible and reliable results. For the LHC, a good knowledge of the aperture restrictions has been one of the key ingredients to push its performance in terms of peak luminosity over the years. In this respect, we stress once more the striking achievement of operating with a \bs which is a factor of two smaller that the design value,  made possible to a large extent by the intense efforts of devising new techniques to probe the beam aperture as well as good optics correction, control of beam-beam effects and implementation of novel optics.

Among all the methods developed, the controlled white noise emittance blow-up with collimator scans became the standard approach to aperture measurements in the last years of the LHC Run~2, due to the better control of the beam losses generated by blowing-up individual bunches. In addition, the collimator scan procedure is much safer, starting with the machine fully protected by the collimator jaws. Furthermore, it allows a simpler aperture reconstruction analysis, as it does not require additional orbit or optics measurements. This enabled us to carry out aperture measurements at 6.5~TeV on a regular basis, which was inconceivable before this approach was devised and demonstrated. The method proved to be reliable and reproducible over the years.

For the design phase of a particle accelerator, the $n_1$ model for aperture calculations, which incorporates the information about tolerances on the key beam and mechanical parameters, has to be used. However, it was shown that in order to get a good agreement with measurements, better assumed tolerances are needed. A realistic, beam-based aperture model, including a set of tolerances based on measurements, is hence of paramount importance for every high-performance particle accelerator. Through such an effort the typical uncertainty in the LHC model has been confirmed to be 0.5~\sig. It is also worth stressing that the combined use of aperture measurements and model, in what has been called the scaling method, proved to be a reliable means to predict an aperture bottleneck starting from a given measurement under different optical configurations. All this was crucial to consolidate the measurement techniques that allowed pushing the performance of the LHC. Such a refined aperture model is also an excellent investment for the future, as it can be used to make predictions on future accelerator designs, such as that of the High Luminosity LHC or a Future Circular Collider.
\begin{acknowledgement}
We would like to express our warm gratitude to the members of the CERN Operations Group, and the Collimation and the Optics and Measurements Corrections team for the continuous support during the many experimental sessions of beam-based aperture measurement. We would also like to thank Daniel Valuch for many constructive discussions. 
\end{acknowledgement}

\appendix

\section{Detailed aperture measurements summary\label{table_injection} at injection energy (450 GeV)}
\begin{table*}[!htbp]
\begin{center}
\caption{Summary of the measured global aperture bottlenecks per beam and plane at injection energy. The aperture measurements are expressed by the reference collimator half-gap before and after exposing the bottleneck. All values are given in units of beam size \sig, for \en=3.5~$\mu$m. The bottleneck location is also indicated for each beam and plane and the beam-based aperture measurement method used.}

\label{tab:aperture_injection}
\small
\begin{tabular}{@{}ccccccc@{}}
                                 \hline    &\begin{tabular}[c]{@{}c@{}}\textbf{E }\\\textbf{[TeV]} \end{tabular}&                           \textbf{Method}&
                                 \textbf{B1H}                                                                              & \textbf{B1V}                                                                   & \textbf{B2H}                                                        & \textbf{B2V}
                                                                  
                         \\\hline  \midrule
                         
\begin{tabular}[c]{@{}c@{}}2010\end{tabular} &0.450 &\begin{tabular}[c]{@{}c@{}}Tune resonance blow-up\\+coll. scan\end{tabular} &\begin{tabular}[c]{@{}c@{}}12.5-13.0\\(Q6R2)   \end{tabular}                       & \begin{tabular}[c]{@{}c@{}}13.5-14.0\\(Q4L6)   \end{tabular}            & \begin{tabular}[c]{@{}c@{}}14.0-14.5\\(Q5R6)   \end{tabular} & \begin{tabular}[c]{@{}c@{}}13.0-13.5\\(Q4R6)  \end{tabular} \\ \midrule
\begin{tabular}[c]{@{}c@{}}2011\end{tabular}&0.450 &\begin{tabular}[c]{@{}c@{}}Tune resonance blow-up\\+coll. scan \end{tabular} & \begin{tabular}[c]{@{}c@{}}12.0-12.5\\(Q6R2)   \end{tabular}                       & \begin{tabular}[c]{@{}c@{}}13.0-13.5\\(Q4L2)   \end{tabular}             & \begin{tabular}[c]{@{}c@{}}12.5-13.0\\(Q5R6)   \end{tabular} & \begin{tabular}[c]{@{}c@{}}13.1-13.6\\(Q4R6)   \end{tabular} \\ \midrule
\begin{tabular}[c]{@{}c@{}}2012\end{tabular} &0.450 &\begin{tabular}[c]{@{}c@{}}Controlled white noise\\emittance blow-up+coll. scan \end{tabular}& \begin{tabular}[c]{@{}c@{}}11.5-12.0\\(Q6R2)   \end{tabular}                       & \begin{tabular}[c]{@{}c@{}}12.0-12.5\\(Q4L6)   \end{tabular}            & \begin{tabular}[c]{@{}c@{}}12.5-13.0\\(Q5R6)   \end{tabular}    & \begin{tabular}[c]{@{}c@{}}12.5-13.0\\(Q4R6)   \end{tabular} \\ \midrule

\begin{tabular}[c]{@{}c@{}}2015\end{tabular}&0.450 &\begin{tabular}[c]{@{}c@{}}Controlled white noise\\emittance blow-up+coll. scan \end{tabular} & \begin{tabular}[c]{@{}c@{}}11.6-12.1\\(MBRC.4R8)   \end{tabular}                       & \begin{tabular}[c]{@{}c@{}}12.1-12.6\\(Q6L4)  \end{tabular}            & \begin{tabular}[c]{@{}c@{}}12.5-13.0\\(Q6L6)   \end{tabular} & \begin{tabular}[c]{@{}c@{}}12.0-12.5\\(Q4R6)   \end{tabular} \\ \midrule
\begin{tabular}[c]{@{}c@{}}2016\end{tabular}&0.450 &\begin{tabular}[c]{@{}c@{}}Controlled white noise\\emittance blow-up+coll. scan \end{tabular} & \begin{tabular}[c]{@{}c@{}}12.5-13.0\\(MBRC.4R8)   \end{tabular}                        & \begin{tabular}[c]{@{}c@{}}12.0-12.5\\(Q6L4)  \end{tabular}             &  \begin{tabular}[c]{@{}c@{}}12.5-13.0\\(TCDQM.4L6)   \end{tabular}
&  \begin{tabular}[c]{@{}c@{}}12.5-13.0\\(Q4R6)   \end{tabular}
\\ \midrule

\begin{tabular}[c]{@{}c@{}}2017\end{tabular}&0.450 &\begin{tabular}[c]{@{}c@{}}Controlled white noise\\emittance blow-up+coll scan \end{tabular} & \begin{tabular}[c]{@{}c@{}}13.1-13.6\\ (Q6R2) \end{tabular}                       & \begin{tabular}[c]{@{}c@{}}12.2-12.7\\(Q4L6)   \end{tabular}            & \begin{tabular}[c]{@{}c@{}}13.2-13.7\\(Q6L8)   \end{tabular} & \begin{tabular}[c]{@{}c@{}}12.8-13.3\\(Q4R6)   \end{tabular} \\ \midrule
\begin{tabular}[c]{@{}c@{}}2018\end{tabular}&0.450 &\begin{tabular}[c]{@{}c@{}}Controlled white noise\\emittance blow-up+coll scan \end{tabular} &\begin{tabular}[c]{@{}c@{}}13.3-13.8\\(Q4R6)   \end{tabular} & \begin{tabular}[c]{@{}c@{}}12.2-12.7\\(Q4L6) \end{tabular}& \begin{tabular}[c]{@{}c@{}}13-13.5\\ (Q4L6 \&  Q6L8) \end{tabular} & \begin{tabular}[c]{@{}c@{}}12.5-13.0\\ (Q4R6)  \end{tabular} \\  \midrule
\end{tabular}
\end{center}
\end{table*}

\section{Detailed aperture measurements summary\label{table} for physics configuration (6.5 TeV)}
\begin{table*}[!htbp]
\begin{center}
\caption{Summary of measured local aperture bottlenecks per plane in IR1/5 in Run~1.  The aperture measurements are given by the reference collimator half-gap before and after exposing the bottleneck. All values are given in units of beam size \sig, for \en=3.5~$\mu$m.}

\label{tab:aperture_collision_local}
\small
\begin{tabular}{@{}cccccccccc@{}}
                                 \hline    &\begin{tabular}[c]{@{}c@{}}\textbf{E}\\\textbf{[TeV]} \end{tabular}&   \begin{tabular}[c]{@{}c@{}}\textbf{\bs}\\ \textbf{[cm]} \end{tabular}\ &\begin{tabular}[c]{@{}c@{}}\textbf{$\phi$}\\\textbf{[$\mu$rad]} \end{tabular}&                           \textbf{Method}&
                                 \textbf{IR1(H)}                                                                       & \textbf{IR1(V)}                                       & \textbf{IR5(H)}                       & \textbf{IR5(V)}&
                                  \\\hline  \midrule
\begin{tabular}[c]{@{}c@{}}2011\end{tabular} &3.5&150&-120&\begin{tabular}[c]{@{}c@{}}Local bump\\+TCT scan\end{tabular}& \begin{tabular}[c]{@{}c@{}}19.8-20.3\\   \end{tabular}                        & \begin{tabular}[c]{@{}c@{}}18.3-18.8\\  \end{tabular}             & \begin{tabular}[c]{@{}c@{}}19.8-20.3\\\end{tabular}             &                 \begin{tabular}[c]{@{}c@{}}$>$20.3\\   \end{tabular} \\ \midrule   
\begin{tabular}[c]{@{}c@{}}2011\end{tabular} &3.5&60&-120&\begin{tabular}[c]{@{}c@{}}Local bump\\+TCT scan\end{tabular}& \begin{tabular}[c]{@{}c@{}}$>$16.0\\   \end{tabular}                        & \begin{tabular}[c]{@{}c@{}}14.8-15.3\\\end{tabular}             & \begin{tabular}[c]{@{}c@{}}15.3-15.8\\ \end{tabular}             &                 \begin{tabular}[c]{@{}c@{}}$>$16.0\\   \end{tabular}
\\ \bottomrule
\end{tabular}
\end{center}
\end{table*}

\begin{table*}[!htbp]
\begin{center}
\caption{Summary of measured global aperture bottlenecks per beam and plane in Run~~1 and Run~2 for physics configuration. The aperture measurements are given by the reference collimator half-gap before and after exposing the bottleneck. All values are given in units of beam size \sig, for \en=3.5~$\mu$m. The bottleneck location is also indicated for each beam and plane.}
\label{tab:aperture_collision}
\small
\begin{tabular}{@{}ccccccccc@{}}
                                 \hline    &\begin{tabular}[c]{@{}c@{}}\textbf{E }\\\textbf{[TeV]} \end{tabular}&   \begin{tabular}[c]{@{}c@{}}\textbf{\bs}\\ \textbf{[cm]} \end{tabular}\ &\begin{tabular}[c]{@{}c@{}}\textbf{$\phi$}\\\textbf{[$\mu$rad]} \end{tabular}&                           \textbf{Method}&
                                 \textbf{B1H}                                                                              & \textbf{B1V}                                                                   & \textbf{B2H}                                                        & \textbf{B2V}
                                                                  
                         \\\hline  \midrule
                         
\begin{tabular}[c]{@{}c@{}}2012\end{tabular} &4&100&-145&\begin{tabular}[c]{@{}c@{}}Controlled white noise \\emittance blow-up+coll. scan\end{tabular}& \begin{tabular}[c]{@{}c@{}}11.5-12.0\\   (Q2L5)\end{tabular}                        & \begin{tabular}[c]{@{}c@{}}11.0-11.5\\   (D1/Q3L1)\end{tabular}             & \begin{tabular}[c]{@{}c@{}}11.5-12.0\\ (D1/Q3R1)\end{tabular}             &                 \begin{tabular}[c]{@{}c@{}}11.0-11.5\\   (D1/Q3R1)\end{tabular} \\ \midrule
\begin{tabular}[c]{@{}c@{}}2015\end{tabular} &6.5 &80&-145&\begin{tabular}[c]{@{}c@{}}Controlled white noise\\emittance blow-up+coll. scan\end{tabular} &\begin{tabular}[c]{@{}c@{}}18.2-18.7\\   (D1/Q3R5)\end{tabular}                       & \begin{tabular}[c]{@{}c@{}}15.7-16.2\\   (D1/Q3L1)\end{tabular}            & \begin{tabular}[c]{@{}c@{}}16.2-16.7\\   (D1/Q3R1)\end{tabular} & \begin{tabular}[c]{@{}c@{}}15.7-16.2\\   (D1/Q3R1)\end{tabular} \\ \midrule
\begin{tabular}[c]{@{}c@{}}2016\end{tabular}&6.5&40&-185 &\begin{tabular}[c]{@{}c@{}}Controlled white noise\\emittance blow-up+coll. scan\end{tabular} & \begin{tabular}[c]{@{}c@{}}10.6-11.1\\   (D1/Q3R5)\end{tabular}                       & \begin{tabular}[c]{@{}c@{}}9.9-10.4\\   (D1/Q3L1)\end{tabular}             & \begin{tabular}[c]{@{}c@{}}11.5-12.0\\   (D1/Q3R1)\end{tabular} & \begin{tabular}[c]{@{}c@{}}10.4-10.9\\   (D1/Q3R1)\end{tabular} \\ \midrule
\begin{tabular}[c]{@{}c@{}}2017\end{tabular} &6.5&40&+185 &\begin{tabular}[c]{@{}c@{}}Controlled white noise\\emittance blow-up+coll. scan\end{tabular}& \begin{tabular}[c]{@{}c@{}}10.9-11.4\\   (D1/Q3R5)\end{tabular}                       & \begin{tabular}[c]{@{}c@{}}12.0-12.5\\   (D1/Q3L1)\end{tabular}            & \begin{tabular}[c]{@{}c@{}}12.9-13.4\\   (Q2R5)\end{tabular}    & \begin{tabular}[c]{@{}c@{}}11.4-11.9\\   (D1/Q3R1)\end{tabular} \\ \midrule

\begin{tabular}[c]{@{}c@{}}2017\end{tabular} &6.5&40&+150 &\begin{tabular}[c]{@{}c@{}}Controlled white noise\\emittance blow-up+coll. scan\end{tabular}& \begin{tabular}[c]{@{}c@{}}11.5-12.0\\   (D1/Q3R5)\end{tabular}                       & \begin{tabular}[c]{@{}c@{}}12.4-12.9\\   (D1/Q3L1)\end{tabular}            & \begin{tabular}[c]{@{}c@{}}14.0-14.5\\   (Q2R5)\end{tabular}    & \begin{tabular}[c]{@{}c@{}}12.0-12.5\\   (D1/Q3L1)\end{tabular} \\ \midrule
\begin{tabular}[c]{@{}c@{}}2017\end{tabular}&6.5&30&+150 &\begin{tabular}[c]{@{}c@{}}Controlled white noise \\emittance blow-up+coll. scan\end{tabular} & \begin{tabular}[c]{@{}c@{}}10.6-11.1\\     (D1/Q3L1\\   \&  D1/Q3R5)\end{tabular} & \begin{tabular}[c]{@{}c@{}}11.1-11.6\\     (Q2R5 \& \\ D1/Q3L1)\end{tabular} & \begin{tabular}[c]{@{}c@{}}10.9-11.4\\   (D1/Q3R1)\end{tabular} & \begin{tabular}[c]{@{}c@{}}10.5-11.0\\   (D1/Q3R1)\end{tabular} \\ \midrule
\begin{tabular}[c]{@{}c@{}}2018\end{tabular}&6.5&30&+160 &\begin{tabular}[c]{@{}c@{}}Controlled white noise\\emittance blow-up+coll. scan\end{tabular} & \begin{tabular}[c]{@{}c@{}}10.5-11.0\\   (D1/Q3L1)\end{tabular}                       & \begin{tabular}[c]{@{}c@{}}10.5-11.0\\   (D1/Q3L1)\end{tabular}            & \begin{tabular}[c]{@{}c@{}}10.0-10.5\\   (D1/Q3R1)\end{tabular} & \begin{tabular}[c]{@{}c@{}}10.5-11.0\\   (D1/Q3R1)\end{tabular} \\ \midrule
\begin{tabular}[c]{@{}c@{}}2018\end{tabular}&6.5&25&+145 &\begin{tabular}[c]{@{}c@{}}Controlled white noise \\emittance blow-up+coll. scan\end{tabular} & \begin{tabular}[c]{@{}c@{}}9.2-9.7\\   (D1/Q3L1)\end{tabular}                        & \begin{tabular}[c]{@{}c@{}}9.2-9.7\\   (D1/Q3L1)\end{tabular}             & \textgreater{}12                                           & \begin{tabular}[c]{@{}c@{}}10.5-11.0\\   (D1/Q3R1)\end{tabular} \\ \midrule
\end{tabular}
\end{center}
\end{table*}
\newpage

\section{Beam-based aperture measurements techniques summary}
In this appendix, Table~\ref{summary_prop} summarises the different beam-based aperture methods developed in the LHC with their applicability, the speed and year of use at the LHC.
\begin{table}[!htbp]
  \begin{center}
    \caption{Summary of beam-based aperture methods.}
    \begin{tabular}{lcccc}
      \hline
      \textbf{Technique}&   \textbf{Type}& \textbf{Speed}&\textbf{Years used }&\textbf{Phase cycle}\\\hline\hline
    Orbit bump&  Local     & Slow &Since 2010&All\\ 
     Tune resonance blow-up& Global   & Slow&2010-2012 & Injection\\
     + coll. scan&       &  &  &\\ 
      Controlled white noise& Global   & Medium&Since 2012 & All\\
     emittance blow-up+coll. scan&       &  &  &\\ 
     Controlled white noise& Global   & Fast&Since 2015 & All\\
     emittance blow-up+coll. alignment&       &  &  & \\\hline
    \end{tabular}
    \label{summary_prop}
  \end{center}
\end{table}
\newpage
\printbibliography 

@Article{tomas09prstab,
  Title                    = {First $\beta$-beating measurement and optics analysis for the CERN Large Hadron Collider},
  Author                   = {Aiba, M. and Fartoukh, S. and Franchi, A. and Giovannozzi, M. and Kain, V. and Lamont, M. and Tom\'as, R. and Vanbavinckhove, G. and Wenninger, J. and Zimmermann, F. and Calaga, R. and Morita, A.},
  Journal                  = {Phys. Rev. ST Accel. Beams},
  Year                     = {2009},
  Pages                    = {081002},
  Volume                   = {12},

  Issue                    = {8},
  Numpages                 = {8},
  Publisher                = {American Physical Society}
}

@Article{ipac11_assmann_aperture,
  Title                    = {{Aperture Determination in the LHC Based on an Emittance Blowup Technique with Collimator Position Scan}},
  Author                   = {R.W. Assmann and R. Bruce and M. del Carmen Alabau and M. Giovannozzi and G.J. Mueller and S. Redaelli and F. Schmidt and R. Tomas and J. Wenninger and D. Wollmann},
  Journal                  = {Proceedings of IPAC11, San Sebastian, Spain},
  Year                     = {2011},
  Pages                    = {1810},

  Url                      = {http://accelconf.web.cern.ch/AccelConf/IPAC2011/papers/tupz006.pdf}
}

@Article{assmann02dump,
  Title                    = {The Consequences of Abnormal Beam Dump Actions on the {LHC} Collimation System},
  Author                   = {R.W. Assmann and B. Goddard and E. Vossenberg and E. Weisse},
  Journal                  = {LHC Project Note 293, CERN},
  Year                     = {1996}
}

@Article{lhcdesignV1,
  Title                    = {{LHC} design report v.1 : The {LHC} main ring},
  Author                   = {O. S. Br{\"{u}}ning and P.~Collier and P.~Lebrun and S.~Myers and R.~Ostojic and J.~Poole and {P.~Proudlock (editors)}},
  Journal                  = {CERN-2004-003-V1},
  Year                     = {2004}
}

@PhdThesis{chiara-thesis,
  Title                    = {{Commissioning Scenarios and Tests for the LHC Collimation System}},
  Author                   = {C. Bracco},
  School                   = {EPFL Lausanne},
  Year                     = {2008}
}

@Article{bruce10_evian,
  Title                    = {{How low can we go? Getting below $\beta^*$=3.5 m}},
  Author                   = {R. Bruce and R.W. Assmann},
  Journal                  = {Proceedings of the 2010 {LHC} beam operation workshop, Evian, France},
  Year                     = {2010},
  Pages                    = {133},

  Url                      = {http://indico.cern.ch/getFile.py/access?contribId=5&sessionId=3&resId=0&materialId=paper&confId=107310}
}

@Article{bruce13_NIM_backgrounds,
  Title                    = {{Sources of machine-induced background in the ATLAS and CMS detectors at the CERN Large Hadron Collider}},
  Author                   = {R. Bruce and R.W. Assmann and V. Boccone and G. Bregliozzi and H. Burkhardt and F. Cerutti and A. Ferrari and M. Huhtinen and A. Lechner and Y. Levinsen and A. Mereghetti and N.V. Mokhov and I.S. Tropin and V. Vlachoudis},
  Journal                  = {Nucl. Instrum. Methods Phys. Res. A},
  Year                     = {2013},
  Number                   = {0},
  Pages                    = {825 - 840},
  Volume                   = {729},

  Doi                      = {10.1016/j.nima.2013.08.058}
}

@Article{bruce14_PRSTAB_sixtr,
  Title                    = {{Simulations and measurements of beam loss patterns at the CERN Large Hadron Collider}},
  Author                   = {Bruce, R. and Assmann, R. W. and Boccone, V. and Bracco, C. and Brugger, M. and Cauchi, M. and Cerutti, F. and Deboy, D. and Ferrari, A. and Lari, L. and Marsili, A. and Mereghetti, A. and Mirarchi, D. and Quaranta, E. and Redaelli, S. and Robert-Demolaize, G. and Rossi, A. and Salvachua, B. and Skordis, E. and Tambasco, C. and Valentino, G. and Weiler, T. and Vlachoudis, V. and Wollmann, D.},
  Journal                  = {Phys. Rev. ST Accel. Beams},
  Year                     = {2014},

  Month                    = {Aug},
  Pages                    = {081004},
  Volume                   = {17},

  Doi                      = {10.1103/PhysRevSTAB.17.081004},
  Issue                    = {8},
  Publisher                = {American Physical Society},
  Url                      = {http://link.aps.org/doi/10.1103/PhysRevSTAB.17.081004}
}

@Article{bruce15_PRSTAB_betaStar,
  Title                    = {{Calculations of safe collimator settings and ${\ensuremath{\beta}}^{*}$ at the CERN Large Hadron Collider}},
  Author                   = {Bruce, R. and Assmann, R. W. and Redaelli, S.},
  Journal                  = {Phys. Rev. ST Accel. Beams},
  Year                     = {2015},

  Month                    = {Jun},
  Pages                    = {061001},
  Volume                   = {18},

  Doi                      = {10.1103/PhysRevSTAB.18.061001},
  Issue                    = {6},
  Numpages                 = {16},
  Publisher                = {American Physical Society},
  Url                      = {http://link.aps.org/doi/10.1103/PhysRevSTAB.18.061001}
}

@Article{bruce17_NIM_beta40cm,
  Title                    = {Reaching record-low $\beta^*$ at the {CERN Large Hadron Collider} using a novel scheme of collimator settings and optics },
  Author                   = {R. Bruce and C. Bracco and R. De Maria and M. Giovannozzi and A. Mereghetti and D. Mirarchi and S. Redaelli and E. Quaranta and B. Salvachua},
  Journal                  = {Nucl. Instrum. Methods Phys. Res. A},
  Year                     = {2017},

  Month                    = {Jan},
  Pages                    = {19 - 30},
  Volume                   = {848},

  Doi                      = {http://dx.doi.org/10.1016/j.nima.2016.12.039},
  ISSN                     = {0168-9002},
  Url                      = {http://www.sciencedirect.com/science/article/pii/S0168900216313092}
}

@Article{bruce16_inj_aperture_HL,
  Title                    = {{Parameters for aperture calculations at injection for HL-LHC}},
  Author                   = {R. Bruce and C. Bracco and R. De Maria and M. Giovannozzi and S. Redaelli and R. Tomas and F. Velotti and J. Wenninger},
  Journal                  = {CERN-ACC-2016-0328},
  Year                     = {2016},

  Url                      = {https://cds.cern.ch/record/2237427}
}

@Article{bruce19_PRAB_beam-halo_backgrounds_ATLAS,
  Title                    = {Collimation-induced experimental background studies at the {CERN Large Hadron Collider}},
  Author                   = {Bruce, R. and Huhtinen, M. and Manousos, A. and Cerutti, F. and Esposito, L. and Kwee-Hinzmann, R. and Lechner, A. and Mereghetti, A. and Mirarchi, D. and Redaelli, S. and Salvachua, B.},
  Journal                  = {Phys. Rev. Accel. Beams},
  Year                     = {2019},

  Month                    = {Feb},
  Pages                    = {021004},
  Volume                   = {22},

  Doi                      = {10.1103/PhysRevAccelBeams.22.021004},
  Issue                    = {2},
  Numpages                 = {18},
  Publisher                = {American Physical Society},
  Url                      = {https://link.aps.org/doi/10.1103/PhysRevAccelBeams.22.021004}
}

@Article{bruce14_n1_ap_meas,
  Title                    = {Parameters for {HL-LHC} aperture calculations},
  Author                   = {R. Bruce and R. de Maria and S. Fartoukh and M. Giovannozzi and S. Redaelli and R. Tomas and J. Wenninger},
  Journal                  = {CERN Report CERN-ACC-2014-0044},
  Year                     = {2014},

  Url                      = {http://cds.cern.ch/record/1697805?ln=en}
}

@Article{bruce14_evian,
  Title                    = {Collimation and $\beta^*$-reach},
  Author                   = {R. Bruce and S. Redaelli},
  Journal                  = {Proceedings of the 5th Evian Workshop, Evian, France},
  Year                     = {2014},

  Url                      = {http://indico.cern.ch/event/310353/session/1/contribution/4/material/paper/0.pdf}
}

@Article{holzer08a,
  Title                    = {Development, production and testing of 4500 beam loss monitors},
  Author                   = {{E. B.~Holzer \textit{et al}}},
  Journal                  = {Proc. of the European Particle Accelerator Conf. 2008, Genoa, Italy},
  Year                     = {2008},
  Pages                    = {1134}
}

@Article{giovannozzi10lhccwg,
  Title                    = {{Global aperture measurements at 450 GeV with 170 mrad crossing angle}},
  Author                   = {M. Giovannozzi and R. Assmann and R. Giachino and D. Jacquet and L. Ponce and S. Redaelli and J. Wenninger},
  Journal                  = {presentation in the LHC Beam Commissioning Working Group meeting, 2010.09.14.},
  Year                     = {2010}
}

@Article{hermes16_ipac_aperture,
  Title                    = {{Improved Aperture Measurements at the LHC and Results from their Application in 2015}},
  Author                   = {P.D. Hermes and R. Bruce and M. Fiascaris and H. Garcia and M. Giovannozzi and R. Kwee-Hinzmann and A. Mereghetti and D. Mirarchi and E. Quaranta and S. Redaelli and B. Salvachua and G. Valentino},
  Journal                  = {Proceedings of the International Particle Accelerator Conference 2016, Busan, Korea},
  Year                     = {2016},
  Pages                    = {1446},

  Url                      = {http://accelconf.web.cern.ch/AccelConf/ipac2016/papers/tupmw014.pdf}
}

@Article{holzer05,
  Title                    = {{Beam Loss Monitoring System for the LHC}},
  Author                   = {E.B. Holzer and B. Dehning and E. Effinger and J. Emery and G. Ferioli and J.L. Gonzalez and E. Gschwendtner and G. Guaglio and M. Hodgson and D. Kramer and R. Leitner and L. Ponce and V. Prieto and M. Stockner and C. Zamantzas},
  Journal                  = {IEEE Nuclear Science Symposium Conference Record},
  Year                     = {2005},
  Pages                    = {1052},
  Volume                   = {2},

  Doi                      = {10.1109/NSSMIC.2005.1596433}
}

@Article{note111,
  Title                    = {Geometrical acceptance in {LHC} Version 5.0},
  Author                   = {J.B. Jeanneret and R. Ostojic},
  Journal                  = {LHC Project Note 111, CERN},
  Year                     = {1997}
}

@Article{note66,
  Title                    = {Geometrical aperture in {LHC} at injection},
  Author                   = {J.B. Jeanneret and T. Risselada},
  Journal                  = {LHC Project Note 66, CERN},
  Year                     = {1996}
}

@Article{redaelli12_IR2aperture,
  Title                    = {{IR2} aperture measurements at 3.5 {TeV}},
  Author                   = {C. Alabau Pons and A. Arduini and R.W. Assmann and R. Bruce and M. Giovannozzi and J.M. Jowett and E. MacLean and G. Muller and S. Redaelli and R. Tomas and G. Valentino and J. Wenninger},
  Journal                  = {CERN-ATS-Note-2012-017 MD},
  Year                     = {2012},

  Url                      = {http://cdsweb.cern.ch/record/1421274?ln=en#}
}

@Article{assmann05chamonix,
  Title                    = {{Collimators and Beam Absorbers for Cleaning and Machine Protection}},
  Author                   = {{R.W.~Assmann}},
  Journal                  = {Proceedings of the LHC Project Workshop - Chamonix XIV, Chamonix, France},
  Year                     = {2005},
  Pages                    = {261}
}

@Article{assmann06,
  Title                    = {{The Final Collimation System for the {LHC}}},
  Author                   = {{R.W.~Assmann \textit{et al.}}},
  Journal                  = {Proc. of the European Particle Accelerator Conference 2006, Edinburgh, Scotland},
  Year                     = {2006},
  Pages                    = {986}
}

@Article{redaelli11evian,
  Title                    = {{Aperture and optics - measurements and conclusions}},
  Author                   = {S. Redaelli and R. Bruce and X. Buffat and M. Giovannozzi and M. Lamont and G. M\"{u}ller and R. Tomas and J. Wenninger},
  Journal                  = {Proceedings of the 2011 {LHC} beam operation workshop, Evian, France},
  Year                     = {2011},
  Url                      = {https://indico.cern.ch/conferenceOtherViews.py?view=standard&confId=155520}
}

@Article{ipac12_redaelli_aperture,
  Title                    = {Aperture measurements in the {LHC} interaction regions},
  Author                   = {S. Redaelli and C. Alabau Pons and R. Assmann and R. Bruce and M. Giovannozzi and G. Muller and M. Pojer and J. Wenninger},
  Journal                  = {Proceedings of IPAC12, New Orleans, Louisiana, USA},
  Year                     = {2012},
  Pages                    = {508},

  Url                      = {http://accelconf.web.cern.ch/AccelConf/IPAC2012/papers/moppd062.pdf}
}

@PhdThesis{guillaume-thesis,
  Title                    = {Design and Performance Optimization of the {LHC} Collimation System},
  Author                   = {G. Robert-Demolaize},
  School                   = {Universite Joseph Fourier, Grenoble},
  Year                     = {2006}
}

@Article{schmidt06,
  Title                    = {Protection of the {CERN Large Hadron Collider}},
  Author                   = {R Schmidt and R Assmann and E Carlier and B Dehning and R Denz and B Goddard and E B Holzer and V Kain and B Puccio and B Todd and J Uythoven and J Wenninger and M Zerlauth},
  Journal                  = {New Journal of Physics},
  Year                     = {2006},
  Number                   = {11},
  Pages                    = {290},
  Volume                   = {8}
}

@Article{valentino12,
  Title                    = {Semiautomatic beam-based {LHC} collimator alignment},
  Author                   = {Valentino, G. and A\ss{}mann, R. and Bruce, R. and Redaelli, S. and Rossi, A. and Sammut, N. and Wollmann, D.},
  Journal                  = {Phys. Rev. ST Accel. Beams},
  Year                     = {2012},
  Pages                    = {051002},
  Volume                   = {15},

  Issue                    = {5},
  Url                      = {http://prst-ab.aps.org/abstract/PRSTAB/v15/i5/e051002}
}

@Article{valentino17_PRSTAB,
  Title                    = {Final implementation, commissioning, and performance of embedded collimator beam position monitors in the Large Hadron Collider},
  Author                   = {Valentino, G. and Baud, G. and Bruce, R. and Gasior, M. and Mereghetti, A. and Mirarchi, D. and Olexa, J. and Redaelli, S. and Salvachua, S. and Valloni, A. and Wenninger, J.},
  Journal                  = {Phys. Rev. Accel. Beams},
  Year                     = {2017},

  Month                    = {Aug},
  Pages                    = {081002},
  Volume                   = {20},

  Doi                      = {10.1103/PhysRevAccelBeams.20.081002},
  Issue                    = {8},
  Numpages                 = {13},
  Url                      = {https://link.aps.org/doi/10.1103/PhysRevAccelBeams.20.081002}
}

@article{Gabi:Crosstalk,
  title = {Data-driven cross-talk modeling of beam losses in LHC collimators},
  author = {Azzopardi, Gabriella and Salvachua, Belen and Valentino, Gianluca},
  journal = {Phys. Rev. Accel. Beams},
  volume = {22},
  issue = {8},
  pages = {083002},
  numpages = {11},
  year = {2019},
  month = {Aug},
  publisher = {American Physical Society},
  doi = {10.1103/PhysRevAccelBeams.22.083002},
  url = {https://link.aps.org/doi/10.1103/PhysRevAccelBeams.22.083002}
}

@Article{Redaelli:2646800,
      author        = {Hermes, P.D. and others},
      title         = {Simulation Tools for heavy-Ion Tracking Collimation},
      organization  = {CERN},
      publisher     = {CERN},
      address       = {Geneva},
      month         = {December},
      year          = {2018},
      url           = {https://cds.cern.ch/record/2646800},
      doi           = {10.23732/CYRCP-2018-002}
}

@article{Salvachua:2302435,
      author        = "Salvachua, B and Assmann, R W and Bruce, R and Cauchi, M
                       and Deboy, D and Lari, L and Marsili, A and Mirarchi, D and
                       Quaranta, E and Redaelli, S and Rossi, A and Valentino, G",
      title         = "{LHC collimation cleaning and operation outlook}",
      pages         = "155-160. 6 p",
      year          = "2012",
      url           = "https://cds.cern.ch/record/2302435",
}

@article{Daniele:evian19,
  title                    = {Special losses during {LHC} run 2},
  author                   = "Mirarchi, D and Arduini, G and Giovannozzi, M and Lechner, A and Redaelli, S and Wenninger, J",
  journal                  = "Proceedings of the 9th Evian Workshop, Evian, France",
  year                     = "2019",
}

@Article{ULOIPAC,
    title         = {Operation of the LHC with Protons at High Luminosity and High Energy},
      author        = "Papotti, Giulia and Albert, Markus and Alemany-Fernandez, Reyes and Crockford, Guy and Fuchsberger, Kajetan and Giachino, Rossano and Giovannozzi, Massimo and Hemelsoet, Georges-Henry and Höfle, Wolfgang and Jacquet, Delphine and Lamont, Mike and Nisbet, David and Normann, Lasse and Pojer, Mirko and Ponce, Laurette and Redaelli, Stefano and Salvachua, Belen and Solfaroli Camillocci, Matteo and Suykerbuyk, Ronaldus and Uythoven, Jan and Wenninger, Jorg",
      number        = "CERN-ACC-2016-229",
      pages         = "WEOCA01. 4 p",
      year          = "2016",
      url           = "https://cds.cern.ch/record/2207396",
      doi           = "10.18429/JACoW-IPAC2016-WEOCA01",
}

@article{PhysRevAccelBeams.22.061004,
  title = {New approach to LHC optics commissioning for the nonlinear era},
  author = {Maclean, E. H. and Tom\'as, R. and Carlier, F. S. and Camillocci, M. S. and Dilly, J. W. and Coello de Portugal, J. and Fol, E. and Fuchsberger, K. and Garcia-Tabares Valdivieso, A. and Giovannozzi, M. and Hofer, M. and Malina, L. and Persson, T. H. B. and Skowronski, P. K. and Wegscheider, A.},
  journal = {Phys. Rev. Accel. Beams},
  volume = {22},
  issue = {6},
  pages = {061004},
  numpages = {21},
  year = {2019},
  month = {6},
  publisher = {American Physical Society},
  doi = {10.1103/PhysRevAccelBeams.22.061004},
  url = {https://link.aps.org/doi/10.1103/PhysRevAccelBeams.22.061004}
}

@article{Private,
  title = {R. Tom\'as private communication }
}

@article{PhysRevAccelBeams.20.061002,
  title = {LHC optics commissioning: A journey towards 1$\%$ optics control},
  author = {Persson, T. and Carlier, F. and de Portugal, J. Coello and Valdivieso, A. Garcia-Tabares and Langner, A. and Maclean, E. H. and Malina, L. and Skowronski, P. and Salvant, B. and Tom\'as, R. and Bonilla, A. C. Garc\'ia},
  journal = {Phys. Rev. Accel. Beams},
  volume = {20},
  issue = {6},
  pages = {061002},
  numpages = {9},
  year = {2017},
  month = {6},
  publisher = {American Physical Society},
  doi = {10.1103/PhysRevAccelBeams.20.061002},
  url = {https://link.aps.org/doi/10.1103/PhysRevAccelBeams.20.061002}
}
\newpage
\end{document}